\documentclass[acmsmall, nonacm]{acmart}

\AtBeginDocument{%
  \providecommand\BibTeX{{%
    \normalfont B\kern-0.5em{\scshape i\kern-0.25em b}\kern-0.8em\TeX}}}

\renewcommand\footnotetextcopyrightpermission[1]{} 

\usepackage{amsmath}
\usepackage{mathptmx} 
\usepackage{booktabs}
\usepackage{listings}
\usepackage{graphicx}
\usepackage{fancyhdr}
\usepackage{xspace}
\usepackage[normalem]{ulem}
\usepackage{pifont}
\usepackage{caption}
\usepackage{subcaption}

\setcopyright{none}

\begin{document}


\title[SPOTS: An Accelerator for Sparse Convolutional Neural Networks DCS-TR-756]{SPOTS: An Accelerator for Sparse Convolutional Neural Networks
  Leveraging Systolic General Matrix-Matrix Multiplication \\ \small{Rutgers Department of Computer Science Technical Report DCS-TR-756}}

\author{Mohammadreza Soltaniyeh}
\affiliation{%
  \institution{Rutgers University}
  \city{New Brunswick}
  \country{USA}
}
\email{m.soltaniyeh@cs.rutgers.edu}

\author{Richard P. Martin}
\affiliation{%
  \institution{Rutgers University}
  \city{New Brunswick}
  \country{USA}
}
\email{rmartin@scarletmail.rutgers.edu}

\author{Santosh Nagarakatte}
\affiliation{%
  \institution{Rutgers University}
  \city{New Brunswick}
  \country{USA}
}
\email{santosh.nagarakatte@cs.rutgers.edu}

\newcommand{\imcol}{\textsc{Im2Col}\xspace}
\newcommand{\etc}{\emph{etc.}\xspace}
\newcommand{\eg}{\emph{e.g.}\xspace}
\newcommand{\ie}{\emph{i.e.}\xspace}

\newcommand{\cmark}{\ding{51}}%
\newcommand{\xmark}{\ding{55}}%


%


%
\begin{abstract}

This paper proposes a novel hardware accelerator for inference of
sparse convolutional neural networks (CNNs) by building a hardware
unit to perform Image to Column (\imcol) transformation of the input
feature map coupled with a systolic array-based general matrix-matrix
multiplication (GEMM) unit. Our design carefully overlaps the \imcol
transformation with the GEMM computation to maximize parallelism.  We
propose a novel design for the \imcol unit that uses a set of
distributed local memories connected by a ring network, which improves
energy efficiency and latency by streaming the input feature map only
once. The systolic array based GEMM unit in the accelerator can be
dynamically configured as multiple GEMM units with square systolic
arrays or as a single GEMM unit with a tall systolic array. This
dynamic reconfigurability enables effective pipelining of \imcol and
GEMM operations and attains high processing element utilization for a
wide range of CNNs. Further, our accelerator is sparsity-aware,
improving performance and energy efficiency by effectively mapping the
sparse feature maps and weights to the processing elements, skipping
ineffectual operations and unnecessary data movements involving
zeros. Our prototype, SPOTS, is on average 1.74$\times$ and
2.16$\times$ faster than two systolic array-based hardware
accelerators, Eyeriss and Gemmini, respectively. SPOTS is also
78$\times$, and 12$\times$ more energy-efficient when compared to CPU and GPU implementations, respectively.

\end{abstract}



\settopmatter{printfolios=true}
\maketitle

\section{Introduction}

\textbf{Inference tasks on edge devices.} Neural networks are widely
used in numerous domains such as video
processing~\cite{alex:alexnet:2017}, speech
recognition~\cite{Collobert:textprocessing}, and natural language
processing~\cite{vggnet:arxiv,Kaiming:resnet:2016}. They have attained
either near-human or better accuracy with many such tasks. To attain
such accuracy, the training phase involves large datasets with several
weight-update iterations, which can take several hours or even
multiple days to complete. Hence, the training phase is typically
performed in the cloud or on a large cluster of machines.  In contrast
to training, the inference task is performed both in the cloud and at
the edge devices (\eg, mobile devices or in the context of Internet of
Things (IoT)). It is often desirable to compute on the edge devices,
especially when network connectivity is either unavailable or is
limited. The edge devices typically have limited memory and compute
resources with strict requirements on energy usage. Hence, this paper
focuses on designing an efficient hardware accelerator for CNN's
inference task targeting edge devices.

\textbf{Accelerating convolutional neural networks.} Among various
neural networks, convolutional neural networks (CNNs) are widely used
in many applications, such as image processing. CNNs can have multiple
types of layers, including convolution layers, fully connected layers,
and pooling layers, with the majority of the computation belongs to
the convolution layers. A convolution operation involves sliding a
smaller filter window over the input array with a stride size,
producing patches (see Figure~\ref{fig:im2colillustration}). Each CNN
layer has multiple features such as the number of filters, kernel
size, stride size, channel size. This creates a diverse set of layers
with unique features, which makes designing an accelerator that can
perform adequately for all types of CNNs layers challenging. Further,
supporting sparse inputs introduces additional complexity to the
design.

\textbf{Some drawbacks of prior CNN accelerators.} Given the
importance of CNNs in various applications, numerous CNN accelerators
have been explored by the
community~\cite{Gondimalla:sparten:micro:2019,
  Hegde:extensor:micro,Albericio:Cnvlutin,zhang:cambriconx:micro:2016,parashar:scnn:isca:2017,chen:eyeriss:journal:2017,Reagen:Minerva:isca:2016,chen:DaDianNao:micro:2014,han:eie:deep:isca:2016,sharify:laconic:isca:2019,Albericio:Bit-Pragmatic:micro,chen:DaDianNao:micro:2014,sharma:DNNWEAVER:micro:2016,huang:ecnn:micro:2019,Chunhua:GoSPA:isca:2021,gemmini:arxiv:2021}. Many
of these designs are tailored to a particular CNN
architecture~\cite{Xu:reconfig:cnn:TACO:2021,han:eie:deep:isca:2016}. Some
works focus primarily on the convolution
operation~\cite{parashar:scnn:isca:2017,chen:eyeriss:journal:2017},
and thus are inefficient for other CNN layers such as fully connected
or pooling layers. Many works suffer from low hardware resource
utilization for certain layer shape and sizes. Regarding the support
for sparse inputs, many of the prior approaches handle sparsity in
either the
weights~\cite{kung:packingsystolic:asplos:2019,zhang:cambriconx:micro:2016}
or the input feature map~\cite{Albericio:Cnvlutin}. Among the
approaches that support sparsity in both weights and feature maps,
SparTen~\cite{Gondimalla:sparten:micro:2019},
ExTensor~\cite{Hegde:extensor:micro} introduce a high hardware cost to
identify matched elements to multiply among the sparse inputs. Other
approaches such as SCNN~\cite{parashar:scnn:isca:2017} handle sparse
weights and input feature maps but introduce redundant
multiplications.

\textbf{Convolution as matrix-multiplication.}  One approach to
implement CNNs is to realize a convolutional layer as a large, single
General Matrix-Matrix Multiplication (GEMM) using a data
reorganization transformation called \textbf{Image-to-Column
  (\imcol)}. Unsurprisingly, many mainstream frameworks adopt this
approach since highly optimized GEMM primitives are available ~(\eg,
BLAS~\cite{blas} or CuBLAS~\cite{cublas}). One method to accelerate
the convolution computation is to offload the GEMM operation to a
hardware accelerator. However, the \imcol operation accounts for a
reasonable fraction of the execution time (29\% of the total
time). Further, \imcol performs many redundant memory accesses,
contributing to the overall energy consumption due to the high energy
cost of memory accesses. Thus, offloading only the GEMM operation to a
hardware accelerator and doing the \imcol operation in software
prevents fine-grained pipelining of the \imcol transformation and the
matrix-multiplication operation. Further, performing the \imcol
operation in hardware helps to avoid significant data transfer between
the CPU and the hardware accelerator.

\textbf{This paper.} We make a case for building a hardware
accelerator that implements the convolution layer as a single large
GEMM operation using \imcol. Our accelerator for sparse convolutional
networks, which we call SPOTS, performs \imcol in hardware along with
the GEMM operation. It effectively pipelines the \imcol operation with
the GEMM operation and eliminates redundant memory accesses. In
addition, our design supports sparse weights and feature maps tailored
for our GEMM and \imcol pipeline. Finally, we achieve generality by
supporting various CNN layers, such as fully connected and pooling
layers, while maintaining high processing element (PE) utilization for
various CNN layers.

\textbf{A dedicated \imcol unit in SPOTS.} We propose a dedicated
hardware \imcol unit that operates in parallel with the hardware GEMM
unit.  This specialized \imcol unit enables us to avoid redundant
accesses and promote data reuse, significantly accelerating inference
and improving energy consumption.  A novel aspect of the \imcol unit
in SPOTS is that it has a collection of patch units~(PUs) that stream
the input only once, performs data reorganization, creates multiple
patches in parallel, and eliminates redundant accesses.  To eliminate
redundant accesses, each patch unit in the \imcol unit has three local
buffers that identify overlapped elements between patches and avoid
costly DRAM accesses. These patches are subsequently fed into a
systolic array-based GEMM unit.

\textbf{A dynamically reconfigurable GEMM unit in SPOTS.} The GEMM
unit in SPOTS is efficiently pipelined with the \imcol unit. The GEMM
unit in SPOTS can be configured as multiple GEMM units with
square-shaped systolic arrays with processing elements (PEs) or a
single tall-thin unit. The tall-thin shape better balances the memory
bandwidth requirement of the GEMM unit and the throughput of \imcol
unit, which allows efficient pipelining of operations between the PEs
performing the matrix multiplication with the PUs executing the \imcol
reorganization. This dynamic reconfigurability of the GEMM units
enables SPOTS to achieve high PE utilization with various kinds of
convolutional layers that differ in the number of filters, kernel
size, stride values, and feature map dimensions. In addition to the
convolution and fully connected layers, SPOTS support pooling layers
with a minor enhancement to the \imcol unit.

\textbf{SPOTS is sparsity-aware.} SPOTS efficiently handles zeros in
both inputs: weights and the input feature map. Sparsity in weights
results from the pruning step in CNNs.  Pruning reduces computation
and memory footprint by eliminating weights after the training phase
without substantively changing network accuracy. Pruning results in
sparse matrices; that is, portions of the array have many zero
elements. SPOTS exploits sparsity to skip data transfer and
computation for sparse regions. Our new sparse format, tailored for
our group-wise pruning, substantially reduces the storage requirement
for the weights in comparison to random pruning
techniques~\cite{han:eie:deep:isca:2016} and provides high bandwidth
for our tall-thin systolic array. Finally, SPOTS tags blocks of zeros
in the result of the \imcol unit and skips zero elements before
entering the systolic array, saving computation cycles and memory
transfers.

The three key innovations in our accelerator are: (1) a novel \imcol
unit that allows it to pipeline GEMM and \imcol computations to
improve performance, (2) dynamically reconfigurable GEMM unit with the
capability to adapt to different CNN layers and shapes, and (3)
sparsity awareness that allows the design to exploit the sparsity in
both the feature map and filters. These techniques combine to improve
CNN performance and energy efficiency over prior accelerators. We
evaluate our design for four popular CNNs, AlexNet, VGGNet, ResNet,
and GoogleNet, that feature a diverse set of convolution layers with
different memory and computation requirements. We compare the
performance and energy efficiency of SPOTS with other state-of-the-art
hardware accelerators for CNNs. Our results show that SPOTS is on
average 1.74$\times$ and 2.16$\times$ faster than
Eyeriss~\cite{chen:eyeriss:journal:2017} and
Gemmini~\cite{gemmini:arxiv:2021}, respectively. SPOTS is also
78$\times$ and 12$\times$ more energy efficiency when compared to CPU
and GPU systems, respectively.  In addition, we demonstrate that SPOTS
can achieve high PE utilization under different CNN shapes. Finally,
we show that our novel \imcol unit improves the energy efficiency by
60\% compared to an \imcol unit which does not reuse the data.

\section{Background and Motivation}
\label{section:background}

We provide background on CNNs, structuring the convolution operation
as general matrix-matrix multiplication with the help of the
\textsc{\imcol} transformation, and leveraging sparsity in the inputs
to improve performance and energy efficiency.

\subsection{Convolution Neural Networks}

A Convolution Neural Network (CNN) consists of a series of
layers. Each layer in a CNN extracts a high-level feature of the input
data called a \textit {feature map} (fmap).  CNNs often have different
layers, including convolution, activation (e.g., non-linear operator),
pooling, and fully connected layers. The convolutional layers are the
main layers in a CNN. They perform the bulk of the computation. Each
convolution layer has several filters. The values of these filters
(i.e., weights) are learned during the training phase. In the
inference phase, the network classifies new inputs presented to the
network.

\begin{figure}[t]
  \centering
  \includegraphics[width=0.7\textwidth,angle=0]{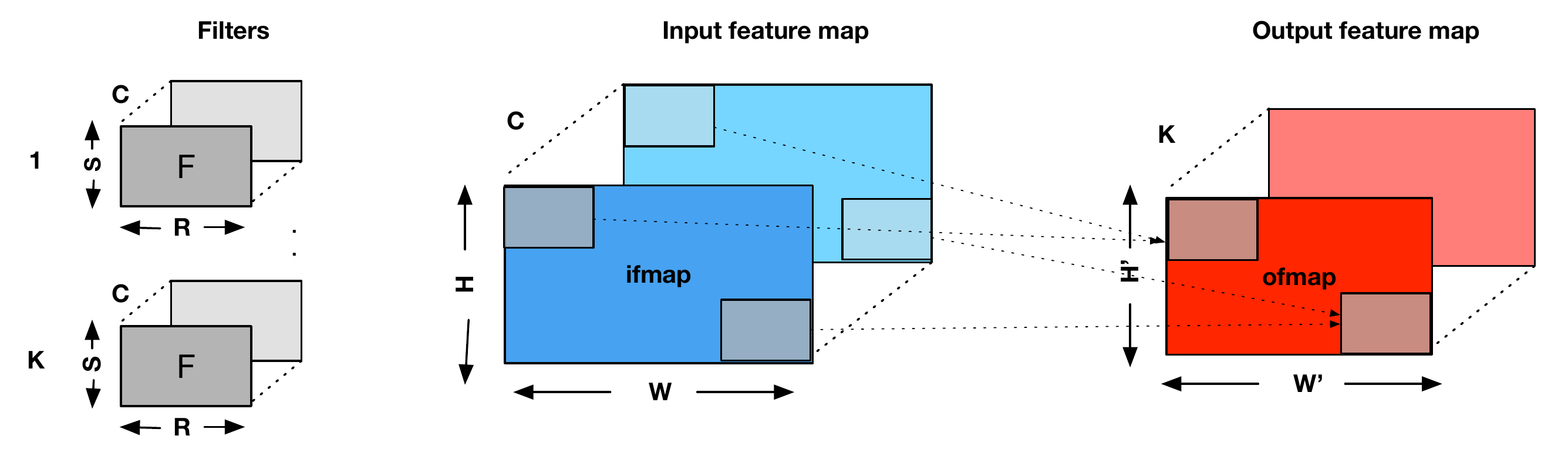}
\caption{\small Illustration of a convolution layer along with its
  inputs.}
\label{fig:convlayer}

\vspace{-4pt}
\end{figure}

Figure~\ref{fig:convlayer} visualizes the computation in the
convolution layer. The input feature map is structured as a 3-D tensor
with W, H, and C as its width, height, and the number of channels,
respectively. Similarly, the filters are structured as 3-D tensors
with width (R), height (S), and C channels. The filters and the input
feature maps have the same number of channels. There are K filters in
this example. Typically, a collection of N input feature maps are
convolved with K filters (\ie, a batch size of N). For inference
tasks, it is common to use a batch size of 1. For some convolution
layers, a 1-D scalar bias is also added to the result, which is not
shown in Figure~\ref{fig:convlayer}.

\textbf{One method to build a hardware accelerator for CNNs.}  The
sliding-window nature of the convolution operation introduces overlaps
between the patches. It makes the job of designing the hardware
accelerator challenging because mapping the computation of a
convolution operation to a set of processing elements (PEs) is more
complex. One commonly used method is to design a fetch unit within
each PE that fetches the input patch, communicates the patches with other PEs, and
manages the partial results. A specialized interconnect is typically
used to facilitate the communication between the PEs based on the specific dataflow.
Prior work such as SCNN~\cite{parashar:scnn:isca:2017} and
Eyeriss~\cite{chen:eyeriss:journal:2017} adopt this approach.  The
main weakness of this approach is that the interconnection network and
dataflow are heavily customized for the convolution operation. Hence,
both SCNN and Eyeriss can be inefficient for other layers, such as
fully connected layers. For example, SCNN can achieve 25\% of the peak
throughput when used for the fully connected layers. Similarly,
Eyeriss fails to achieve high PE utilization for small batch sizes.

\subsection{Transforming Convolution to General Matrix-Matrix
  Multiplication}
  \label{back-im2col}

\begin{figure}[t]
  \centering
  \includegraphics[width=0.6\textwidth,angle=0]{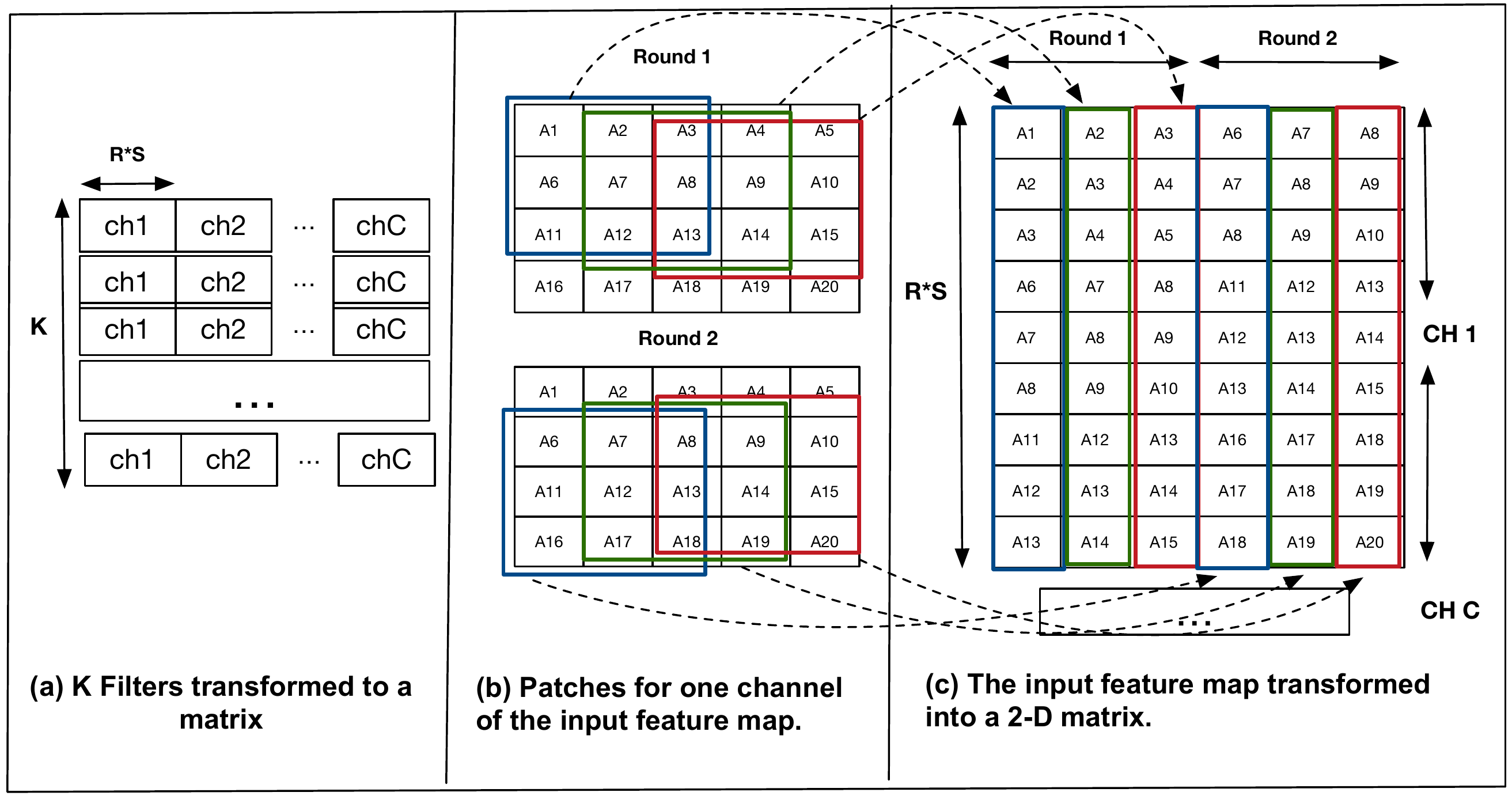}
\caption{\small Transforming the inputs of a convolution layer (i.e.,
  input feature map and filters) into two matrices to use a GEMM-based
  formulation of convolution.}
\label{fig:im2colillustration}
\end{figure}

The convolution operation can be transformed into general
matrix-matrix multiplication (GEMM) using the \imcol transformation.
To structure the convolution operation as matrix multiplication, we
need to create two matrices from the two inputs of a convolution layer:
input feature map and the K filters.
Figure~\ref{fig:im2colillustration} illustrates how the two
matrices are built. The product of these two matrices will be equivalent
to the result of the convolution operation.
For building the weight matrix, each filter is mapped to one row of
the weight matrix. When there are K filters, there will be K rows in
the weight matrix~(see Figure~\ref{fig:im2colillustration}(a)). The
number of columns in the weight matrix is $R\times S \times C$.
In contrast to the weight matrix, a more complex transformation is
required to build a 2-D matrix from the original 3-D input feature
map. This transformation is called \textbf{Image to Column
  (\imcol)}. The \imcol result depends on the kernel size and the
stride size, which are the two parameters of the convolution
operation.  In convolution, each filter slides across different
positions in the input feature map. We call all elements in the input
feature map covered by the filter as a \textit{patch} or a tile.
Patches are often overlapped with each other when the stride size is
less than the filter size. This overlap results in the repetition of
the same element of the input feature map in multiple
patches. Figure~\ref{fig:im2colillustration}(b) and
Figure~\ref{fig:im2colillustration}(c) illustrates the \imcol
transformation with an example filter of size ($3 \times 3 \times C$)
and a stride of 1. Each column of the matrix produced by the \imcol
transformation corresponds to one patch where the filter is applied
for all $C$ channels, and it has $R \times S \times C$
rows. Figure~\ref{fig:im2colillustration} shows the patches for one
channel. Finally, the product of the two matrices
(Figure~\ref{fig:im2colillustration}(a)
and~\ref{fig:im2colillustration}(c)) generates the output of the
convolution operation.

\begin{figure}[t]
  \centering
  \includegraphics[width=0.9\textwidth,angle=0]{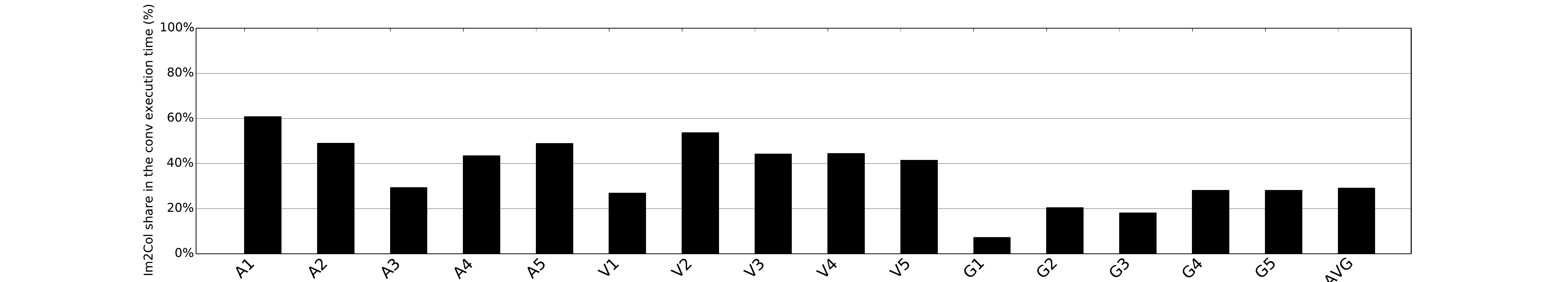}
\caption{\small The percentage of the total execution time spent in
  the \imcol transformation for various convolution layers from
  AlexNet, VGGNet, and GoogleNet for a CPU system.}
\label{fig:im2colperc}

\end{figure}

\textbf{Benefits and challenges of convolution with \imcol.}  By using
a separate \imcol transformation, the task of building input patches
and the eventual computation on them can be decoupled. The \imcol
transformation can identify data overlap among different patches as
each filter slides across different positions in the input feature
map. Further, a separate \imcol transformation can enable one to use highly optimized primitives or even available hardware accelerators for GEMM. However, doing the \imcol
transformation in software may not provide the best possible
performance because of redundant accesses with the GEMM computation and increased data storage/communication.  
The \imcol transformation can account for as much as 60\% of the total execution time of the
convolution operation. Figure~\ref{fig:im2colperc} reports the
percentage of the total execution time that is spent in the \imcol
transformation for various layers in AlexNet, VGGNet, and GoogleNet
for a CPU system.
On average, the \imcol transformation spends 29\% of the overall
execution time. Additionally, a naive \imcol transformation can result
in numerous redundant memory accesses. Several repetitions in the
\imcol patches are created by sliding the filters over the input
feature map. Depending on the filter size and the stride size, the
number of memory access can be $9\times$ higher on average than the
number of elements, which indicates that many elements are redundantly
accessed multiple times.

Section~\ref{sec:spots} describes our accelerator that performs \imcol
on-the-fly, extracts significant parallelism between various patches,
and use the hardware \imcol unit to simplify the hardware accelerator
for GEMM without the need for complex interconnection networks.

\begin{figure}[t]
  \centering
  \includegraphics[width=0.7\textwidth,angle=0]{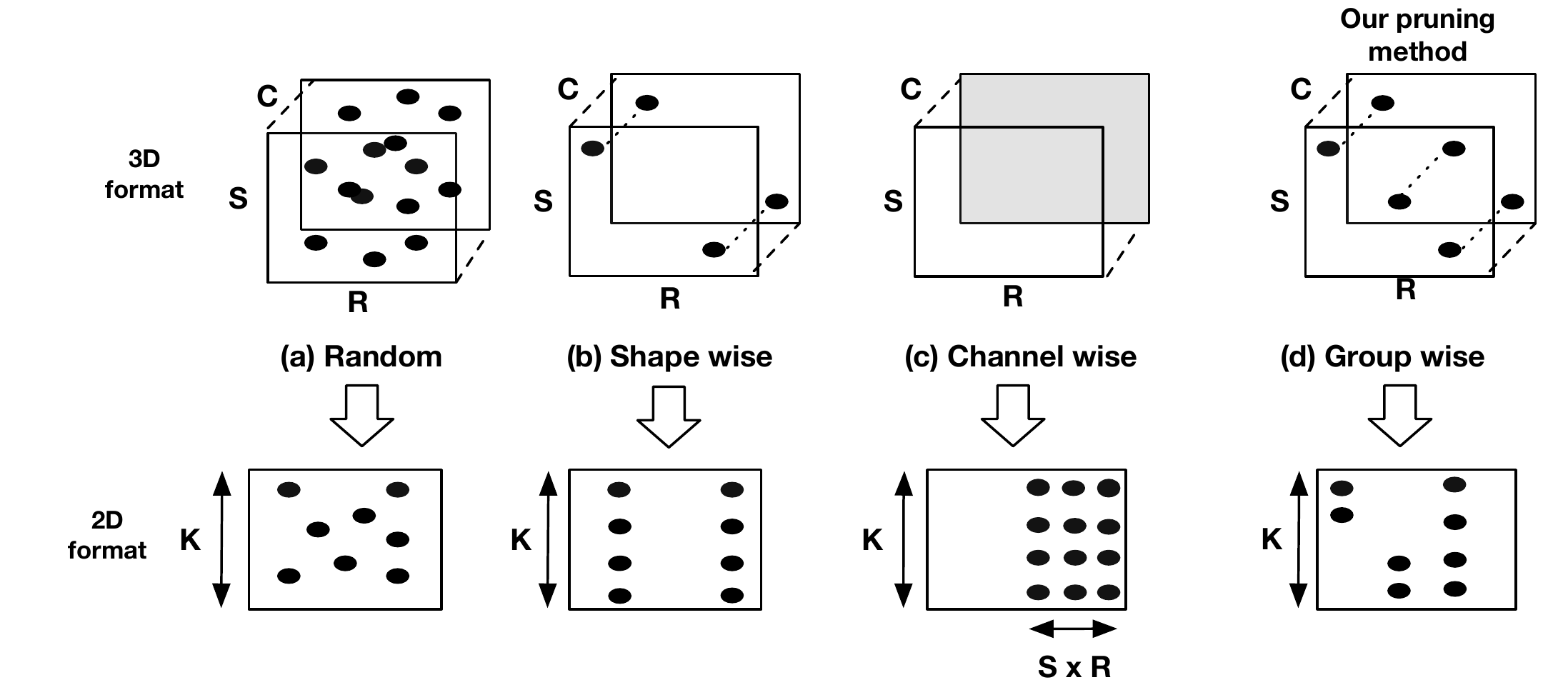}
\caption{\small This figure shows the resulting zeros in the 2-D
  matrix representation of the filter while pruning the filters at
  different granularities and their corresponding matrix format. A
  dark dot indicates that the point is being pruned.}
\label{fig:pruningformat}
\end{figure}

\vspace{-1.2em}

\subsection{Sparsity-Awareness in CNNs} 
\label{back-sparseaware}
A fraction of the values in the layers' weight and feature map are
zeros.  During training, a pruning step is often applied to remove
unimportant and redundant weights, which can result in numerous zeros
in the final trained weights. Additionally, some zeros can also appear
in the feature map input. Unlike zeros in the weights, these zeros
need to be identified at run-time.

The pruned weights can be compressed using a sparse format. In
addition to reducing the model size, different hardware accelerators
use sparsity at different levels to improve their design's performance
and energy efficiency. The performance improvement comes from eliding
multiplications and minimizing data movement when it involves
zeros. Next, we review some of the sparsity-awareness techniques and
their weaknesses to motivate our proposal.

\textbf{Techniques for skipping the zeros.}
SparTen~\cite{Gondimalla:sparten:micro:2019} and
ExTensor~\cite{Hegde:extensor:micro} add extra hardware to find the
matched non-zero values to multiply (i.e., intersect operation). The
hardware cost of the intersection operation is high (\eg, a prefix sum
for SparTen and content addressable memory for
ExTensor). SCNN~\cite{parashar:scnn:isca:2017} adopts another approach
to avoid zeros in both feature maps and weights by using an outer
product. While the SCNN approach successfully removes the costly
intersection operation, it introduces redundant multiplications in an effort to handle overlaps in the input tiles.

Unlike these approaches, we avoid the zeros in the input
controller. It relieves the PEs from doing extra costly operations
(e.g., intersection) and redundant operations.  We will explain the
details of our sparsity-awareness design in
Section~\ref{sec:sparsity-aware}.


\textbf{Techniques for pruning filters.} There are two strategies for
pruning: random pruning and structured pruning. The \emph{random
pruning} sets a weight to zero if it is below a threshold
value~\cite{han:deepcompression:arxiv:2016}. Typically after the
pruning step, non-zero weights need to be stored in a compressed
sparse format. However, using a sparse format involves indirect
accesses and requires extra steps for extracting the non-zero elements
and matching indices. In contrast, \emph{structured pruning} address
the irregular accesses due to random
pruning~\cite{Wei:ssl:nips:2016,kung:packingsystolic:asplos:2019,Kang:accelertorprune:2019}.
Structured pruning removes redundant weights only at well-defined
locations or with specific block sizes.

Figure~\ref{fig:pruningformat} shows pruning at different levels with
various pruning methods. The dark points represent pruned weights in
the filter. When we convert a 3-D filter to a 2-D representation using
the strategy shown in Figure~\ref{fig:im2colillustration}(a),
the resulting zeros in the 2-D matrix is shown in the second row of
Figure~\ref{fig:pruningformat}. Random pruning results in an irregular
pattern of zeros. A coarse-grained structure (\eg, channel-wise) for
pruning can result in a group of zero columns in the 2-D matrix, which
is more hardware friendly. However, it can
sacrifice network accuracy. A fine-grained structure (e.g.,
shape-wise or group-wise) gets closer to the accuracy of random pruning while having
a regular structure with zeros. We will describe the details of our group-wise pruning in Section~\ref{section:methodology}.

\section{SPOTS Architecture}
\label{sec:spots}

\begin{figure}[t]
		
        \begin{subfigure}[b]{0.51\textwidth}
        \centering
                \includegraphics[width=\textwidth]{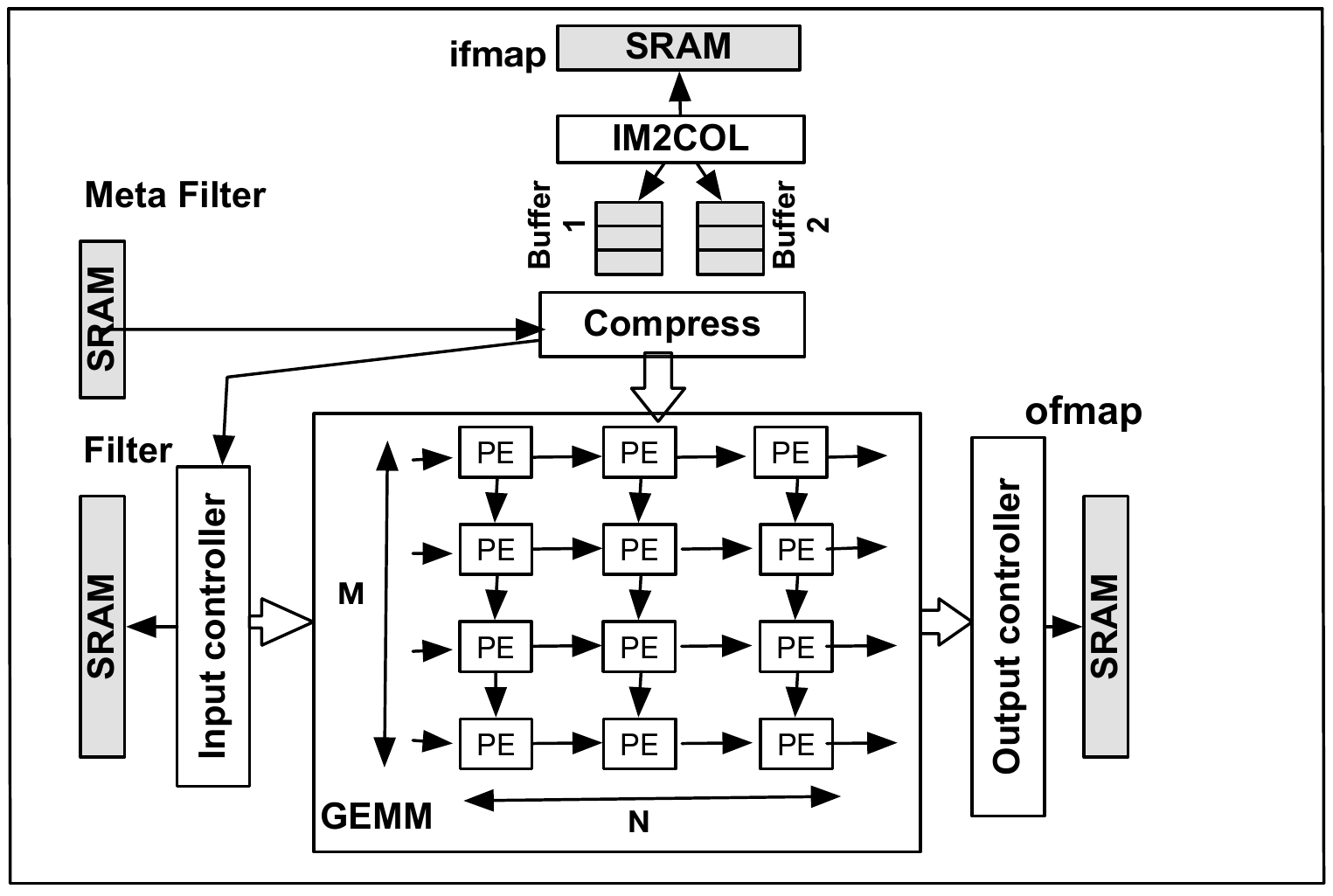}
                \caption{\small Overall architecture of SPOTS}
                \label{fig:overallarch}
              
        \end{subfigure}%
        \begin{subfigure}[b]{0.49\textwidth}
        \centering
                \includegraphics[width=\textwidth]{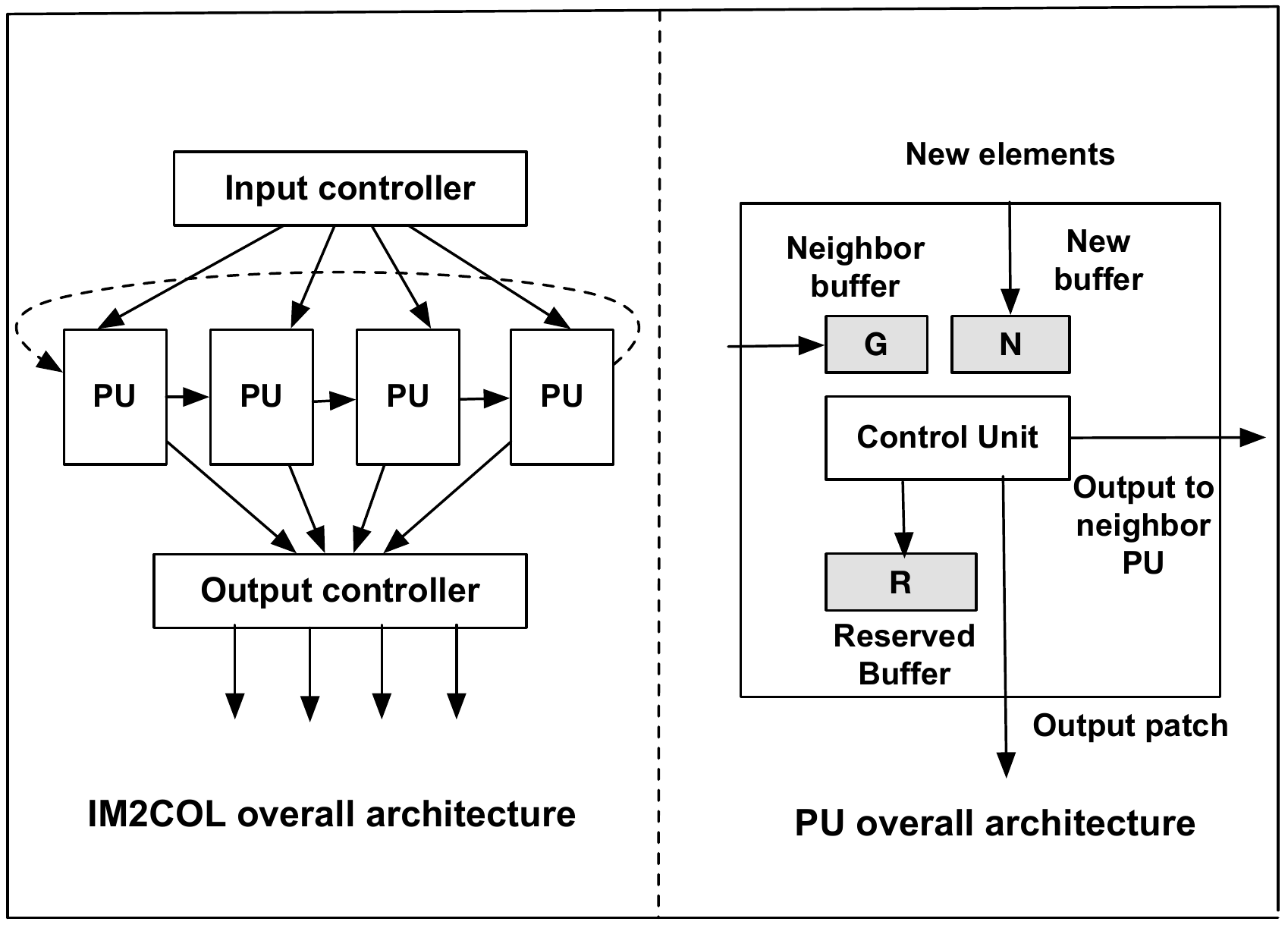}
                \caption{\small Overall \imcol architecture and patch unit internals.}
                \label{fig:im2colarch}
     
        \end{subfigure}%
        
        \caption{\small (a) The overall architecture of our accelerator with the \imcol  unit and a systolic array-based GEMM unit. (b) The Overall \imcol architecture and patch unit internals. }
        
\end{figure}

We design a hardware accelerator, SPOTS, for the inference phase that
provides significant performance and energy benefits for CNNs with
different layer characteristics using a GEMM-based formulation of a
convolution operation. Our design goals are four-fold: (1) significant
performance and energy benefits, (2) support multiple CNN layers and
filters of varying sizes, (3) efficient even with sparsity in the
weights and the filters, and (4) fine-grained pipelining of the \imcol
operation with the GEMM computation.

We propose a hardware unit for the \imcol transformation that is
synergistic and pipelined with the hardware unit for GEMM.  The \imcol
unit reads the input feature map, a 3-D array, and creates a set of
linearized patches. The \imcol unit consists of patch units (PUs)
where each PU is responsible for constructing a linear patch. As
values are streamed in, the PU constructing the patch will forward
overlapped elements to neighboring PUs. Once the PU collects all the
values in a patch, it forwards in-order partial patches to the GEMM
unit. The approach allows the \imcol unit to read in values from the
input feature map once and reuse them to save redundant memory
accesses.

We design a dynamically reconfigurable GEMM unit with a systolic array
based design. It can be configured as a tall array to balance the
work between \imcol and GEMM computation. To maintain a high PE
utilization with CNN layers with varying shapes, the GEMM units can be
configured as small GEMM units~(see Section~\ref{sec:flexible}).  This
dynamic reconfigurability enables our hardware to adapt to CNN layers
with varying dimensions and shapes. Further, it also helps with
sparsity-awareness by enabling our design to detect and skip zeros in
the input feature map~(see Section~\ref{sec:sparsity-aware}).
Figure~\ref{fig:overallarch} shows the overall architecture of our
accelerator. The two main components are the unit for the \imcol
transformation and the GEMM unit. They are connected by two buffers
that allow effective pipelining of the operations between the \imcol
unit and the GEMM unit.

\subsection{The \imcol Unit}
The \imcol transformation creates a 2-D matrix from the 3-D input
feature map, which reduces convolution to matrix multiplication (See
Section~\ref{back-im2col}). The \imcol transformation is challenging
because it inherits a part of the complexity of convolution, has
complex memory access patterns, and results in redundant accesses.
We propose a distributed hardware structure consisting of a series of
Patch Units (PUs) to both accelerate \imcol and minimize the number of
accesses to the elements of the input feature map. The key insight in
our \imcol unit is to exploit the localities resulting from the
overlap between the patches as we slide the filters across the input
feature map both vertically and horizontally. Each PU is responsible
for building one patch at a time. One of our design goals is to read
the input feature map only once from SRAM. To accomplish this goal,
each patch unit has small local buffers that store some values that
will be useful for building future patches. The PUs are also connected
using a ring network, which allows the PUs to communicate elements
locally and avoid redundant accesses to the input feature map in SRAM.
Figure~\ref{fig:im2colarch} shows the overall architecture of our
\imcol unit that consists of three main components: input controller,
PUs, and output controller.
The input controller reads the input feature map from SRAM and
forwards them to the appropriate PU units.  Apart from sending values
from the input feature map to the respective PUs, the input controller
maintains extra metadata for every scheduled patch. This
metadata carries information about the position of the current
patch. For some convolution layers, stride size is the same as
kernel size. In those cases, there is no overlap between the
patches. For those scenarios, the input control forwards its output
directly to the output controller by skipping the PUs.

Our \imcol unit has multiple PUs within it. The PUs are the main
components of the \imcol unit for generating patches.
Figure~\ref{fig:im2colarch} shows the internals of the PU. Each PU
has three buffers: the new buffer, the neighbor buffer, and the
reserved buffer. The new buffer (N) maintains the newly fetched
element received from the input controller. The neighbor buffer (G)
stores the elements received from the neighboring PU. The reserved
buffer (R) stores some of the elements previously received at that PU
in the previous rounds.  We store the row and column indices (i.e.,
coordinates) along with the value for each element. The control unit
within each PU manages the buffer and generates patches. It decides
whether an element needs to be forwarded to the neighboring PU and
whether it should be maintained in the reserve buffer for future use.

A unique identifier identifies each patch (i.e., row and column
index of top-left element). The control unit in a PU uses the patch
identifier, the filter size, and the stride size to determine which
elements need to be (1) fetched from the input feature map, (2)
forwarded to the neighboring PUs, and (3) stored in the reserve buffer
for future rounds. For example, all elements need to be fetched from
the input feature map when a PU processes the first patch in the first
round.

All elements that are necessary for adjacent patches in a given round
are provided by the neighboring PUs. A PU typically receives $K^2 -
K\times S$ elements from the neighboring patches as long as it is not
the first patch in a given round, where $K$ is the size of the kernel
and $S$ is the stride size.  We assign all patches that belong to the
same column (i.e., column index of the top-left element) in different
rounds to the same PU.  Hence, the PUs also stores some elements that
may be useful to build patches in subsequent rounds in the reserved
buffer. This procedure is repeated for all C channels in the feature
map.

The total number of elements that are overlapped between the vertical
patches for a given filter size is $C \times W \times \left( K-S \right)$ where $W$
is the width of the input feature map. This is the maximum data reuse
that can be attained with the reserve buffer. Further, the width and
the channel size are inversely proportional to each other. For
example, the first few layers of a CNN often have a small number of
channels that are wider.  In contrast, the later layers of the CNN
have larger channels of smaller width. Thus, a small reserve buffer
can provide significant data reuse even for larger layers.  When the
number of overlapped elements between the vertical patches is larger
than the size of the reserved buffer, the input controller skips the
reserved buffer and fetches the element again from SRAM. In such
cases, data reuse is restricted to horizontally adjacent patches.
Finally, the output controller organizes patches formed by each PU and
manages communications with the GEMM unit. It coordinates double
buffering that enables the overlapped execution of the \imcol unit and
the GEMM unit.

\begin{figure}[t]
  \centering
  \includegraphics[width=0.7\textwidth,angle=0]{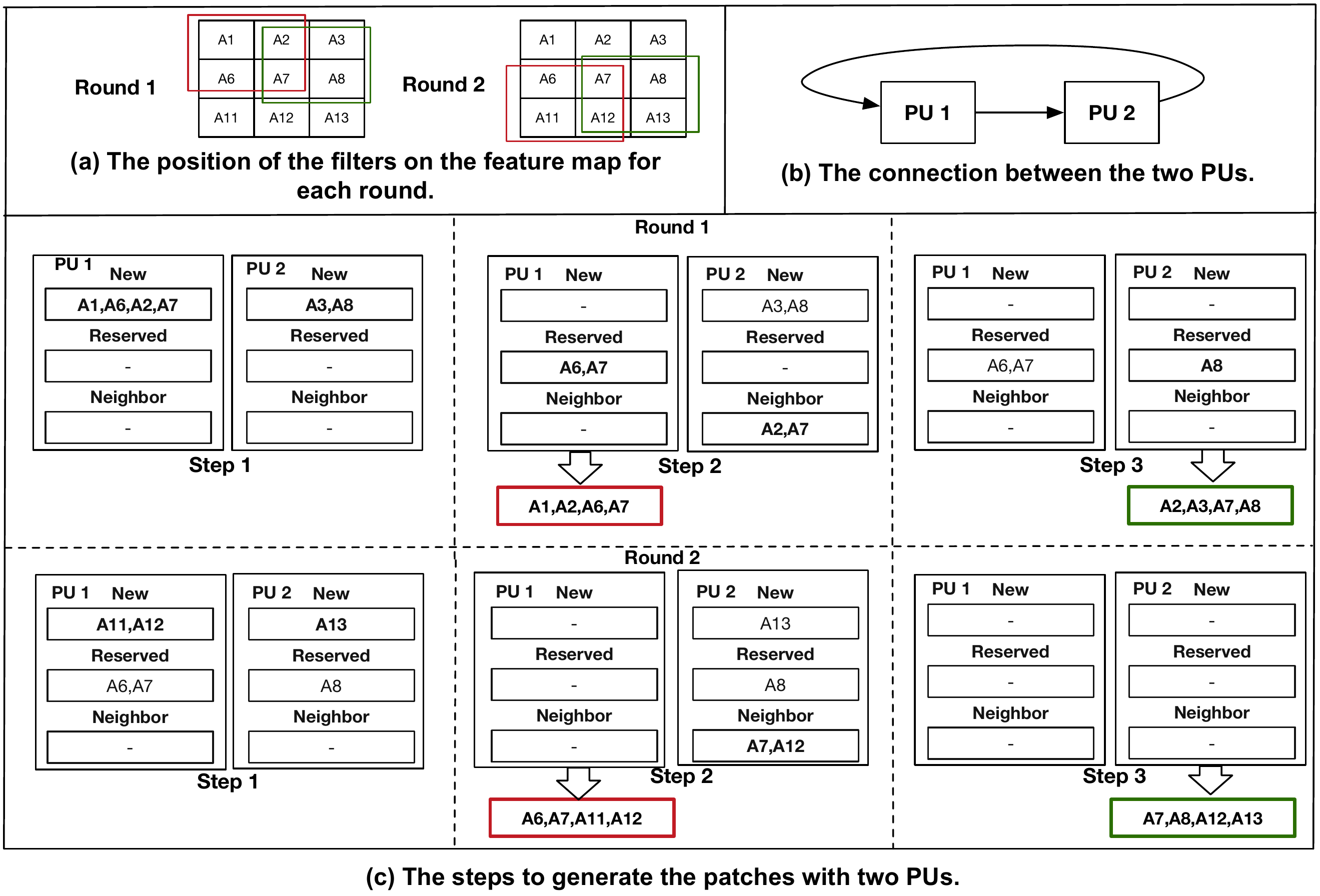}
\caption{\small Illustration of patch generation using the PUs in the
  \imcol unit. We show an \imcol unit with 2 PUs for exposition. (a)
  The input feature map with one channel. We show the sliding windows
  used to generate patches with a stride of 1. (b) The two PUs are
  interconnected by a ring network. (c) There are two rounds. Round 1
  corresponds to patches belonging to the first row of sliding windows
  over the input feature map. Similarly, round 2 corresponds to
  patches belonging to the second row of sliding windows.}
\label{fig:patchgeneration}
\end{figure}

Figure~\ref{fig:patchgeneration} illustrates the process of generating
the patches using the PUs in our \imcol unit. For example, PU1
receives four elements (A1, A6, A2, A7) from the input controller and
stores them in the new buffer in step 1. Similarly, PU2 receives two
new elements (A3, A8). PU2 will receive other elements from the PU1
in subsequent steps (i.e., step 2).

In summary, our hardware \imcol unit provides two benefits: energy
efficiency and performance. Accessing the smaller SRAM and performing
integer operations (for computing on row and column indices) consumes
significantly less energy than accessing DRAM and large SRAMs. Hence,
our design provides significant energy benefits. Further, our
distributed collection of PUs unlocks extra parallelism beyond
parallelism among the channels, allowing multiple patches to be built
simultaneously by different PUs in the \imcol unit that boosts
performance.

\subsection{The GEMM Unit}
\begin{figure}[t]
  \centering
  \includegraphics[width=0.7\textwidth,angle=0]{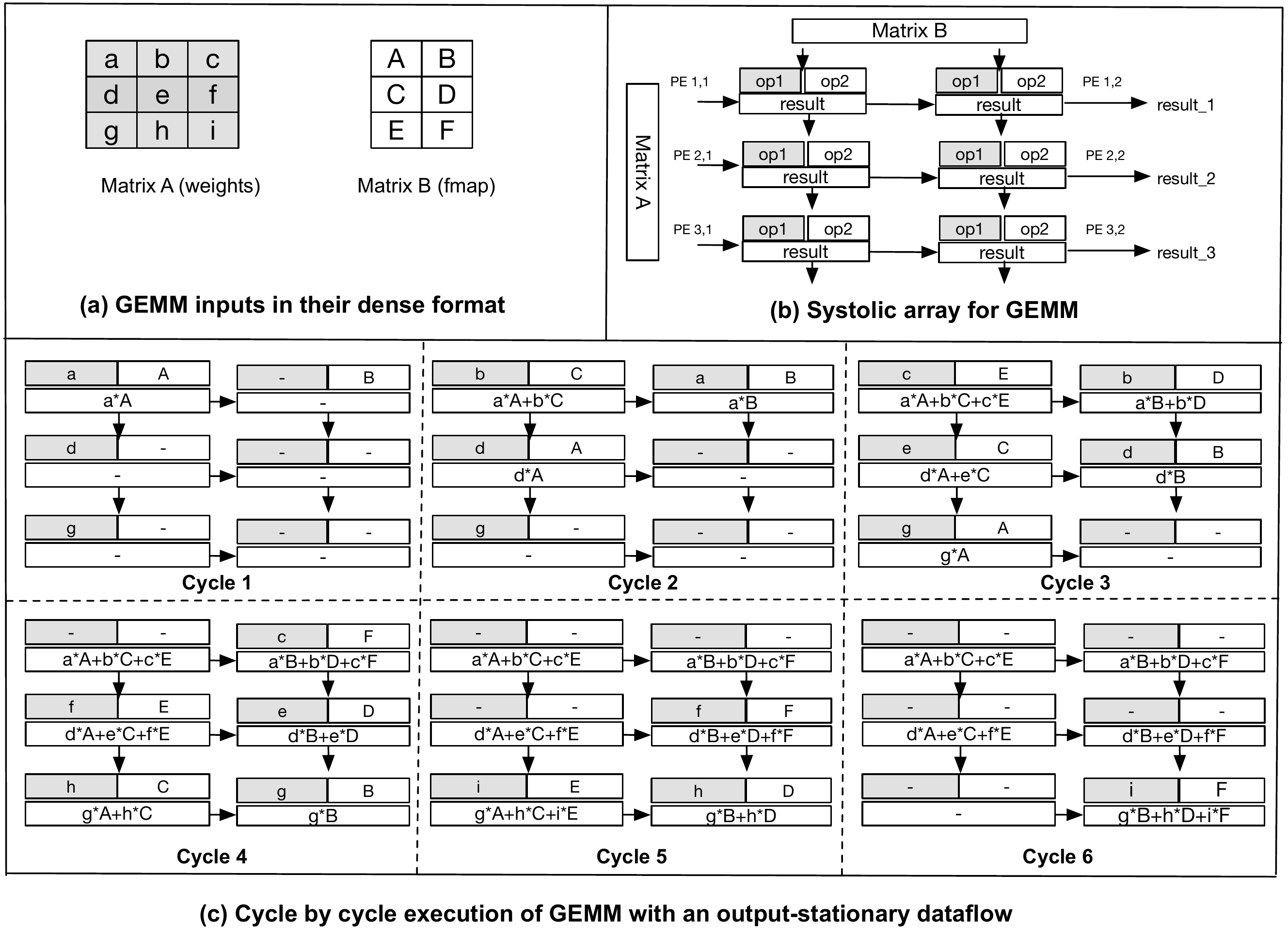}
\caption{\small Illustration of our GEMM unit. (a) Inputs to the GEMM
  unit. (b) A tall array for the GEMM unit. (c) Illustration of GEMM
  computation at various steps. We show the current inputs and the
  partial results computed till a step for each PE. We demonstrate the
  output-stationary attribute of our design.}
\label{fig:gemm}
\end{figure}

Our hardware unit for accelerating GEMM is a systolic array-based
design. Unlike many prior proposals that use systolic arrays for
GEMM~\cite{chen:eyeriss:journal:2017,kung:packingsystolic:asplos:2019,tpu:google:isca,
  kwon:hda:hpca:2019,kung:packingsystolic:asplos:2019}, we add dynamic
reconfigurability to the GEMM unit. The GEMM unit in SPOTS can be
configured either as a tall shaped systolic array (the height is
considerably larger than the width) to maximize data reuse or as
multiple GEMM units with square shaped systolic
arrays. Figure~\ref{fig:gemm}(b) shows our systolic array-based design
for GEMM with a tall array.

There are two main benefits in using a tall-shape systolic array-based
architecture for GEMM. First, one of the inputs of the GEMM unit comes
from the \imcol unit. Using a tall shape array reduces the memory
bandwidth requirement for the input arriving from the \imcol
unit. Thus, we can attain high PE utilization in the GEMM unit with
less throughput from the \imcol unit. This helps us to build an \imcol
unit with fewer resources and memory bandwidth requirements. Second,
the tall array helps our design to exploit sparsity in the output of
the \imcol unit to skip zeros and increase performance.
As the width of the tall array is smaller than its height, fewer
columns from the \imcol transformation enter the systolic array at any
instant of time, which increases the opportunity for detecting and
skipping entire rows of inputs with zeros before entering the systolic
array. In essence, using a tall-shape array helps to simplify our
mechanism to skip the redundant computation involving zeros in the
input feature map. We will explain our sparsity-awareness approach
later (See Section~\ref{sec:sparsity-aware}). Dynamic
reconfigurability enables it to be reorganized as multiple GEMM units
for a few instances where a tall-shape array is not suitable to keep
all the PEs utilized (see Section~\ref{sec:flexible}).

Our GEMM unit uses an output-stationary dataflow where a given
processing element (PE) computes the final result by accumulating the
partial products for a particular element of the output. This
output-stationary dataflow ensures maximum reuse of the output
data. Using a tall array also helps us attain high data reuse for the
result of the \imcol transformation.  Figure~\ref{fig:gemm}(a) shows
the weight matrix from the filter and the output of the \imcol
transformation that forms the input to the GEMM unit. The values of
the filter matrix enter the GEMM unit's systolic array from
left-to-right. While the result of the \imcol unit enters the systolic
array from top-to-bottom. Figure~\ref{fig:gemm}(c) shows the various
steps and partial results computed in the GEMM unit.  Our design is
parameterizable with $M$ rows and $N$ columns in the systolic array.
In our design, each row handles multiple rows of the filter
matrix. Our specific prototype used 128 rows of PEs and 4
columns. These numbers are chosen based on the characteristic of
common CNN layers. Further, each row of the systolic array can be
assigned multiple rows of the filter matrix depending on the
scheduling mode.  The majority of layers in state-of-the-art CNNs have
less than 512 rows of the filter matrix in each convolution layer.

Each PE has a single multiply-accumulate (MAC) unit that uses two
16-bit fixed-point inputs and accumulates the result in a 24-bit
register. To handle multiple rows of the filter matrix, each PE has
$K$ registers to compute the final result (e.g., in our design, we use
$K=4$).  Each PE has three FIFOs.  Two FIFOs, one for each arriving
inputs. The other FIFO works as the work queue for the MAC unit.  In
GEMM, the coordinates of the elements of the two input matrices should
match before multiplying the inputs. In the fetch unit, we ensure that
the inputs are sent to the PEs in the proper order; thus, we do
\textbf{not} need additional logic to perform index matching inside a
PE. Additionally, our output-stationary dataflow ensures all the
partial products produced in a PE belongs to the same output
element. Next, we describe how to support sparsities in both inputs
without requiring any index matching units inside the PEs.

\subsection{Handling Sparsity in CNNs}
\label{sec:sparsity-aware}
Most CNNs have sparsity in both filters and the input feature
map. Figure~\ref{fig:sparsity} quantifies the amount of sparsity
(percentage of the zeros in the total number of elements) for the
commonly used CNNs. We use structured sparsity learning
(SSL)~\cite{Wei:ssl:nips:2016} as our pruning method that is further
enhanced with optimizations to better suit our hardware design (see
Section~\ref{section:methodology}). To support sparsity during
inference, we propose a custom sparse format to store the filters
pruned with our method and design an approach that identifies a block
of entries with all zeros in the result of the \imcol transformation
on-the-fly.  These techniques enable our accelerator to skip rows and
columns with all zeros before entering the systolic array of the GEMM
unit without requiring extra costly hardware for intersection or
introducing any redundant zeros (See
Section~\ref{back-sparseaware}). Further, they also allow us to gate
the MAC units when an operand is zero.  As our designs use a tall
systolic array and an output-stationary dataflow, these techniques
provide high bandwidth access to the filters necessary to keep the PEs
active.

\begin{figure}[t]
  \centering
  \includegraphics[width=0.8\textwidth,angle=0]{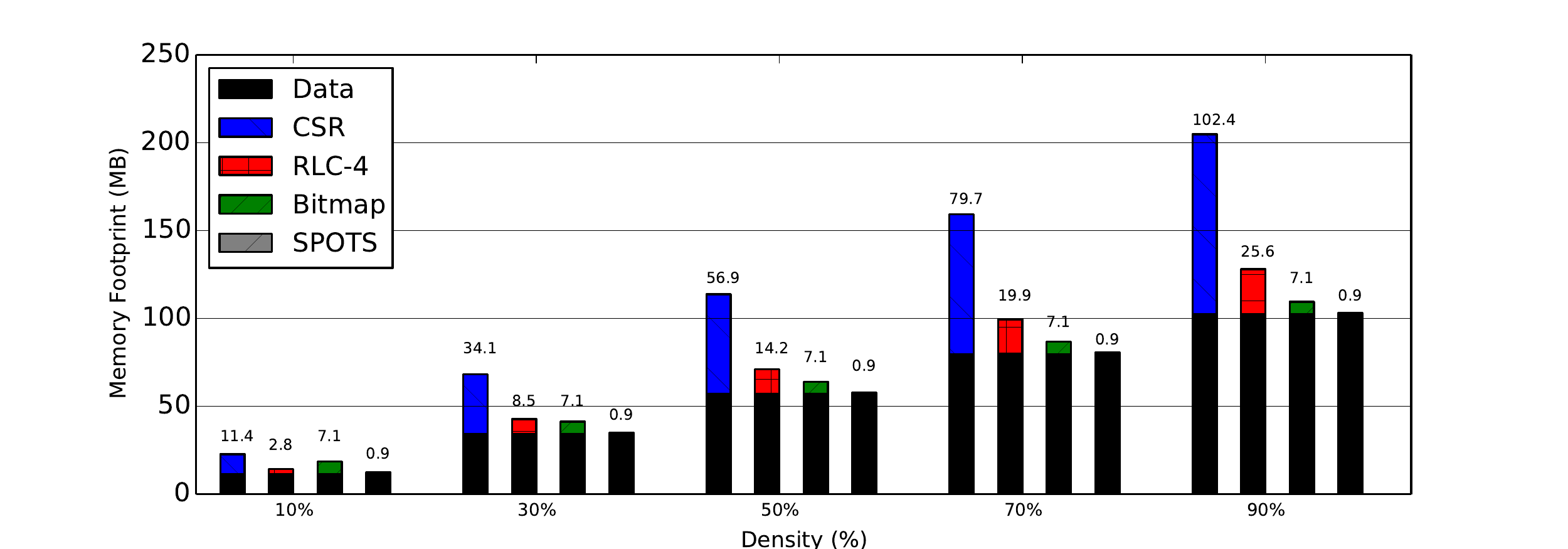}
\caption{\small Comparing our custom sparse format with other
state-of-the-art sparse formats. We used a matrix with 1632 rows and
  36548 columns. We assume the values are 2 bytes. The last bar shows our sparse format. Our sparse format is independent of the density of the zeros in the matrix, and the size of metadata is less than 1 MB across all the density ratios. We compare our sparse format with CSR, RLC-4, and
  Bitmap in the following order. The size of the sparse metadata is shown at the top of each bar in Megabytes.}
\label{fig:sparsecompare}
\end{figure}

\begin{figure}[t]
    
        \begin{subfigure}[b]{0.49\textwidth}
        \centering
                \includegraphics[width=\textwidth]{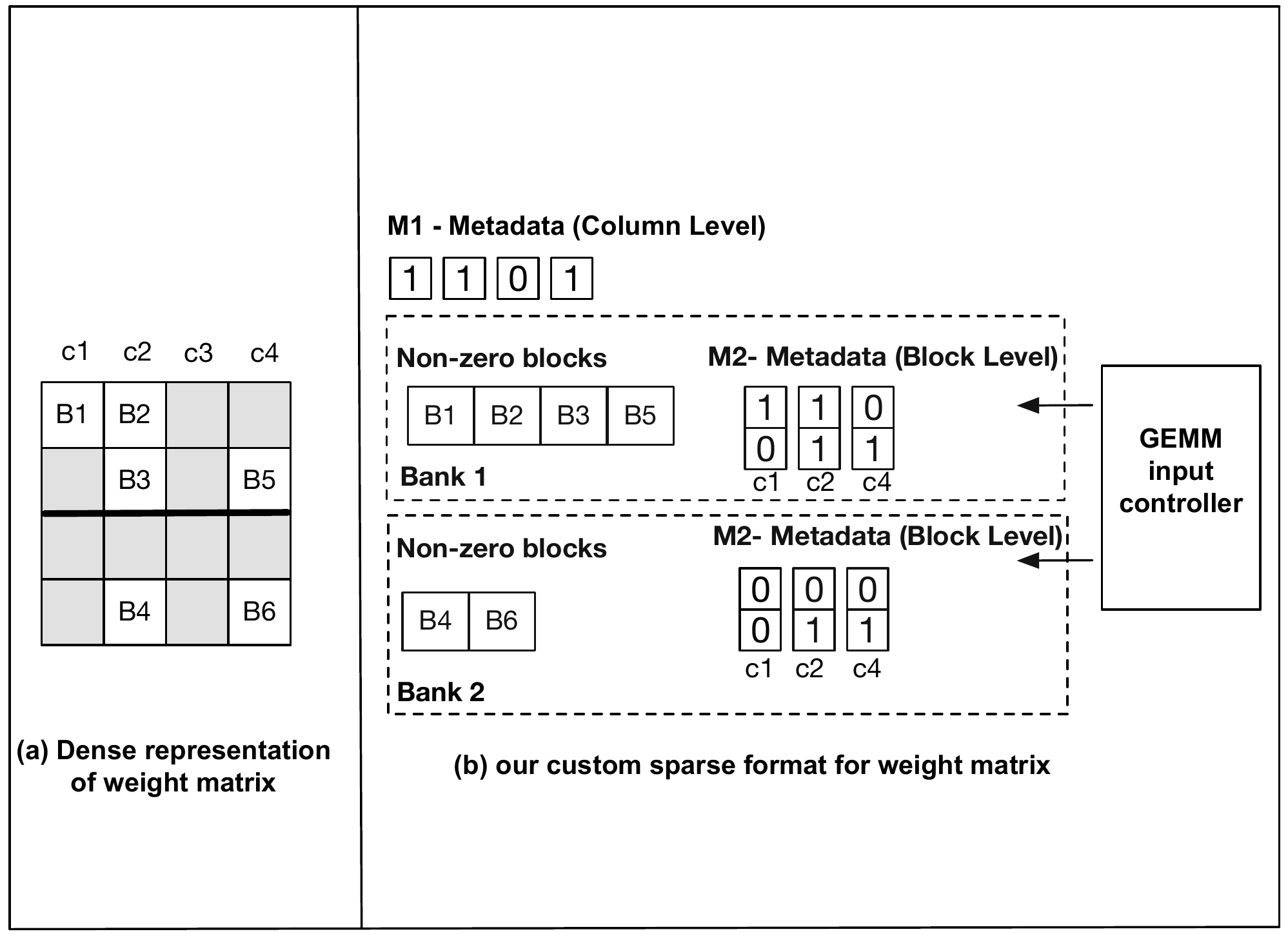}
                
                \caption{\small Our custom sparse format to store filters}
                \label{fig:sparseform}
              
        \end{subfigure}%
        \begin{subfigure}[b]{0.49\textwidth}
        \centering
                \includegraphics[width=\textwidth]{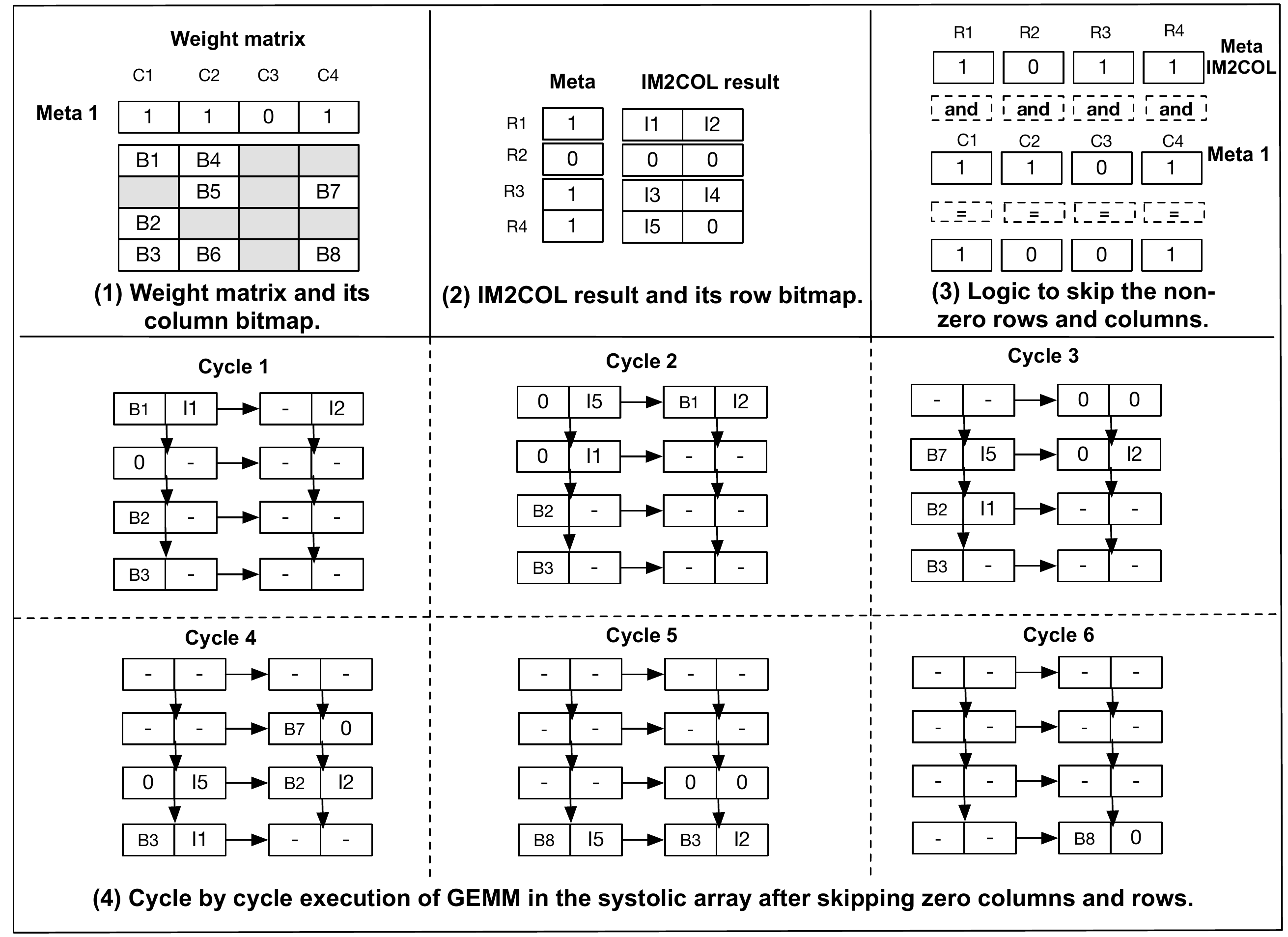}
                
                \caption{\small Skip rows and columns with all zeros}
                \label{fig:skipzero}
     
        \end{subfigure}%
        
        \caption{(a) Our custom sparse format to store filters. (b) Illustration of how our design skips rows and
    columns with all zeros. (1) Weight matrix with the metadata about
    columns with all zeros. (2) The \imcol result with the metadata
    about rows with all zeros. (3) If a row or a column is all zeros,
    all such rows and columns can be skipped (i.e., \textit{and} operation of the row and
    column metadata). (4) GEMM computation when rows and columns are
    skipped. For example, the first element of column C4 will be
    fetched by the first PE in cycle 2 (skipping columns C2 and C3).}
        
\end{figure}

\textbf{Our sparse format for filters.} Once the weights for the
filters are learned during the training phase, we divide the weights
into blocks. The block size is equal to the group size used for
pruning, which is a design parameter. Logically, the filter matrix
will be 2-D matrix of blocks when viewed in the dense
representation. To minimize the memory footprint for storing the filters
during the inference, we convert them into a sparse representation
that is aware of the number of SRAM banks in the design. Our sparse
format uses three arrays to store the pruned weights compactly.
Figure~\ref{fig:sparseform} shows our custom sparse format.  We store
all non-zero blocks separately in one array (Array A) that is
distributed in multiple banks based on the row index of the block (i.e.,
vertical position in the filter matrix).  We use two bitmap arrays M1
and M2 to store the metadata. The bitmap array M1 encodes whether a
column has any non-zeros in the filter matrix. A zero in the bitmap
array M1 indicates an empty column. The bitmap array M2 maintains
whether a block in a non-zero column is non-zero. A zero in M2
indicates the corresponding block is zero (i.e., as a block is a
collection of values, it implies that all values in the block are
zeros).  These three arrays of our sparse format (i.e., A, M1, and M2)
are distributed across the various banks of the SRAM so that the input
controller for the GEMM unit can access them in parallel.
Figure~\ref{fig:sparsecompare} compares the memory footprint of our
sparse format in contrast to traditional compressed sparse row format
and other sparse formats used in prior work. In contrast to them, our
sparse format reduces the memory footprint significantly.

\textbf{Handling sparsity in the result of the \imcol transformation.}
The \textit{compress} component before the GEMM unit in our
accelerator (see Figure~\ref{fig:overallarch}) identifies a block
of zeros in the result of the \imcol transformation. It creates a
bitmap for every block coming out of the \imcol unit. If all elements
in a block in the output of the \imcol unit are zeros, the bit is set
to zero for that block; otherwise, the bit set to one. Subsequently,
the input controller of the GEMM unit uses this bitmap to skip blocks
with all zeros on-the-fly. We can elide MAC operations when an operand
is zero even before entering the systolic array. Further, it is not
necessary to stream the column of filters when one detects such a
block of zeros. Figure~\ref{fig:skipzero} illustrates how the zero
columns in the weight matrix and the zero rows in the output of the
\imcol unit are skipped. The $\star$ marker in
Figure~\ref{fig:sparsity} indicates the percentage of zeros in the
output of the \imcol transformation that is skipped on-the-fly with
this technique. We further reduce energy by gating the MAC units when
an operand is zero.

\subsection{Handling Various CNN Layers/Shapes}
\label{sec:flexible}
\begin{figure}[th]
  \centering
  \includegraphics[width=0.7\textwidth,angle=0]{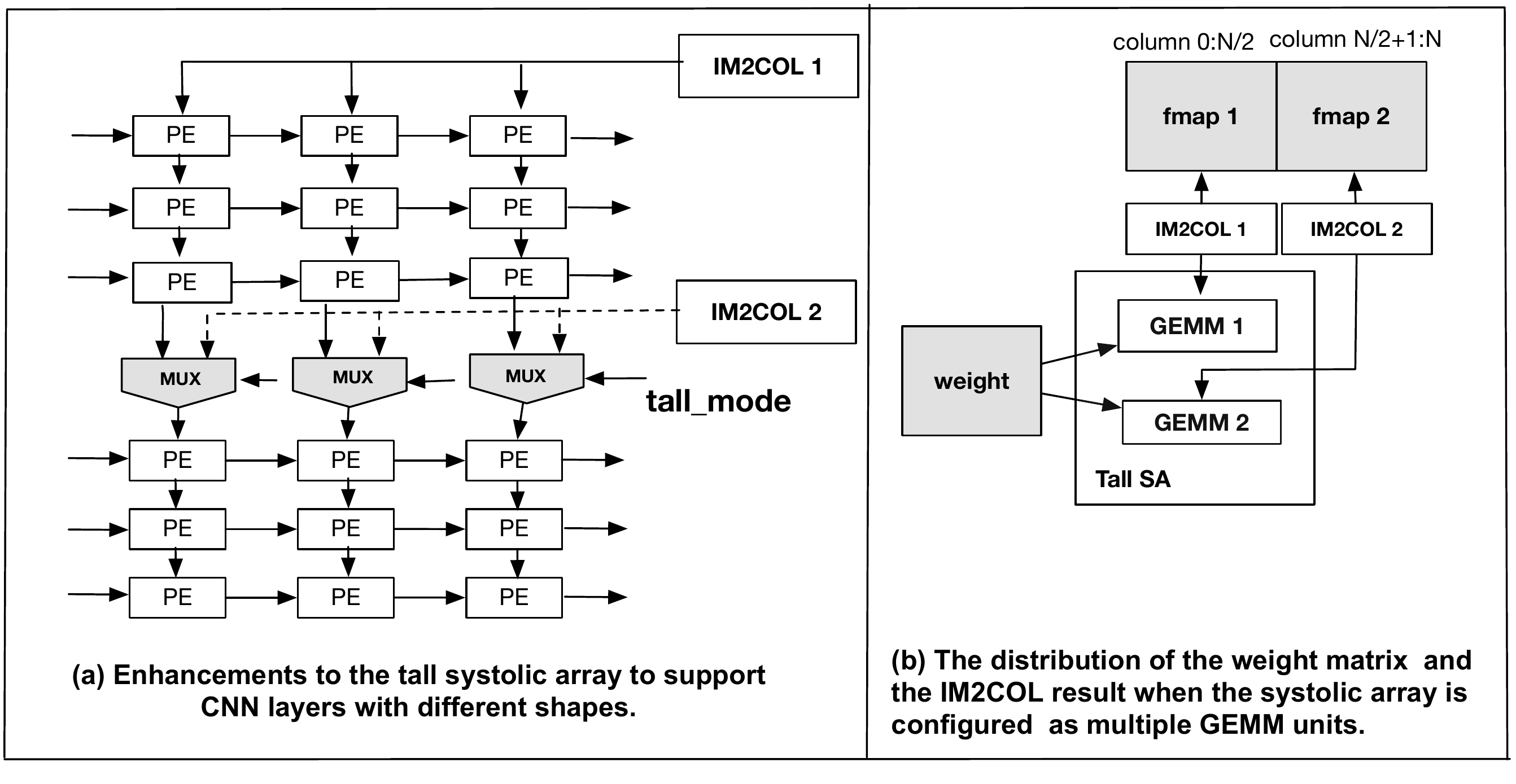}
\caption{\small (a) Enhancements to reorganize the tall systolic array (SA)
  as multiple GEMM units. (b) Illustration of how inputs are
  distributed in the configuration with multiple GEMM units.}
\label{fig:flexibledesign}
\end{figure}

CNNs have multiple layers that can be of different shapes and
sizes. With a fixed configuration of hardware PEs, they can be
underutilized for some layers, shapes and/or sizes.  Each filter forms
a row of the weight matrix that is assigned to a distinct row of the
systolic array. When the GEMM unit is configured as a tall systolic array,
and the number of filters is relatively smaller than the systolic
array's height (e.g., 128), some PEs will remain unused.

Dynamic reconfigurability of the GEMM unit enables us to support CNN
layers with different attributes~(see
Figure~\ref{fig:flexibledesign}). Specifically, the PEs in the GEMM
unit can be configured either as one tall array or multiple small
arrays. Each such configuration has the same number of columns.  This
enhancement allows our design to be more \textbf{adaptive} to
different layer shapes and thus maintains high PE utilization under
different conditions. Figure~\ref{fig:flexibledesign} (a) demonstrates
how a tall array can be used as two smaller arrays using the
multiplexers.
Hence, the PEs now can receive the input either from the PEs above
(i.e., it forms a tall array) or can get the input from a different
\imcol unit.
These multiplexers can be configured based on the mode register
dynamically depending on the structure of a layer. The weight matrix
is broadcast to all small systolic arrays when the GEMM unit is
configured as smaller systolic arrays. Each small GEMM unit receives
the feature map input from their assigned \imcol units. The two GEMM
units compute two independent groups of columns of the final result
matrix (i.e., GEMM 1 computes result columns from 0 to N/2, GEMM
computes the columns from N/2+1 to N). In our prototype, we have four
\imcol units. The main \imcol unit is used when the GEMM unit is used
in the tall array configuration. Other \imcol units are smaller in
size to reduce the overall area. This dynamic reorganization of the
GEMM unit's systolic array coupled with the multiple \imcol units
enables our hardware to maintain high PE utilization for various CNN
layers with different shapes.

\textbf{Supporting fully connected layers.} Most CNNs have one or more
\textit{fully connected} layers at the end of the network. The inputs
to the fully connected layers are the matrix weights learned during
the inference and the output feature map resulting from the final
pooling or convolutional layer that is flattened to a vector.  With a
batch size of 1, the computation for a fully connected layer is
equivalent to matrix-vector multiplication. By increasing the batch
size, we can structure it as a matrix-matrix multiplication
operation. As we use a tall array, the batch sizes need not be large to utilize the whole array of PEs fully
(\eg, can be as small as 4).

\textbf{Supporting pooling layers.} The pooling layers help to
summarize the features generated by a convolution layer. There are two
common types of pooling layers: max pooling and average pooling. Among
them, max pooling, which picks the maximum element from a feature
covered by the filter, is more common. Similar to convolution layers,
the pooling layer has two parameters, filter size and the stride
size. We support the pooling layer by adding the pooling operation
(e.g., MAX) to the output of the patch units (PUs) in the \imcol unit.

\section{EXPERIMENTAL METHODOLOGY}
\label{section:methodology}

\begin{table}[t]
  \centering
   \caption{SPOTS design parameters and area.}.
  \begin{tabular}{|c|c|c|c|} \hline
    Unit &  & Size & Area (mm2)\\ \hline

    & \#PE units & 512 &\\\cline{2-3}
    &Multiplier width & 16 bits & \\\cline{2-3}
    GEMM & Accumulator width & 24 bits & 2.048 \\\cline{2-3}
    & Systolic array & one (128$\times$4) & \\
    & configurations & four (32$\times$4) & \\\cline{2-3}
    &PE's local buffers & 2 KB &\\\hline

    & \#PU units & 4 &\\\cline{2-3}
    \imcol & Reserved buffers  & 32 KB & 1.137\\\cline{2-3} 
    & Other SRAM buffers & 2 MB & \\ \hline
    
    On-chip  & Filter SRAM & 1 MB & 5.426  \\\cline{2-3}
    memory & Fmap SRAM & 512 KB & \\ \hline

    SPOTS total & \multicolumn{2}{|c|}{} & 8.611 \\ \hline
  
  \end{tabular}
 
  \label{table:spotsconfig}
\end{table}

\begin{figure*}[t]
        \begin{subfigure}[b]{0.50\textwidth}
        \centering
                \includegraphics[width=\textwidth]{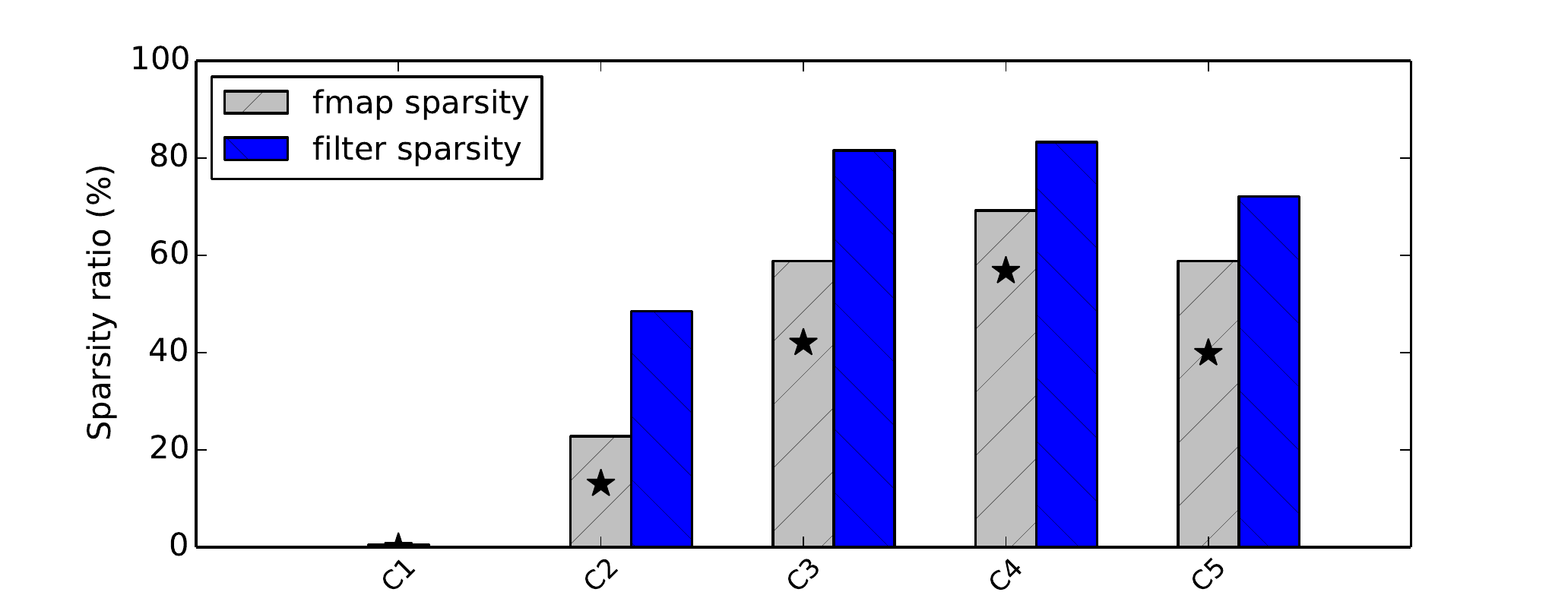}
                \caption{AlexNet.}
              
        \end{subfigure}%
        \hspace{-1.8em}%
        \begin{subfigure}[b]{0.50\textwidth}
        \centering
                \includegraphics[width=\textwidth]{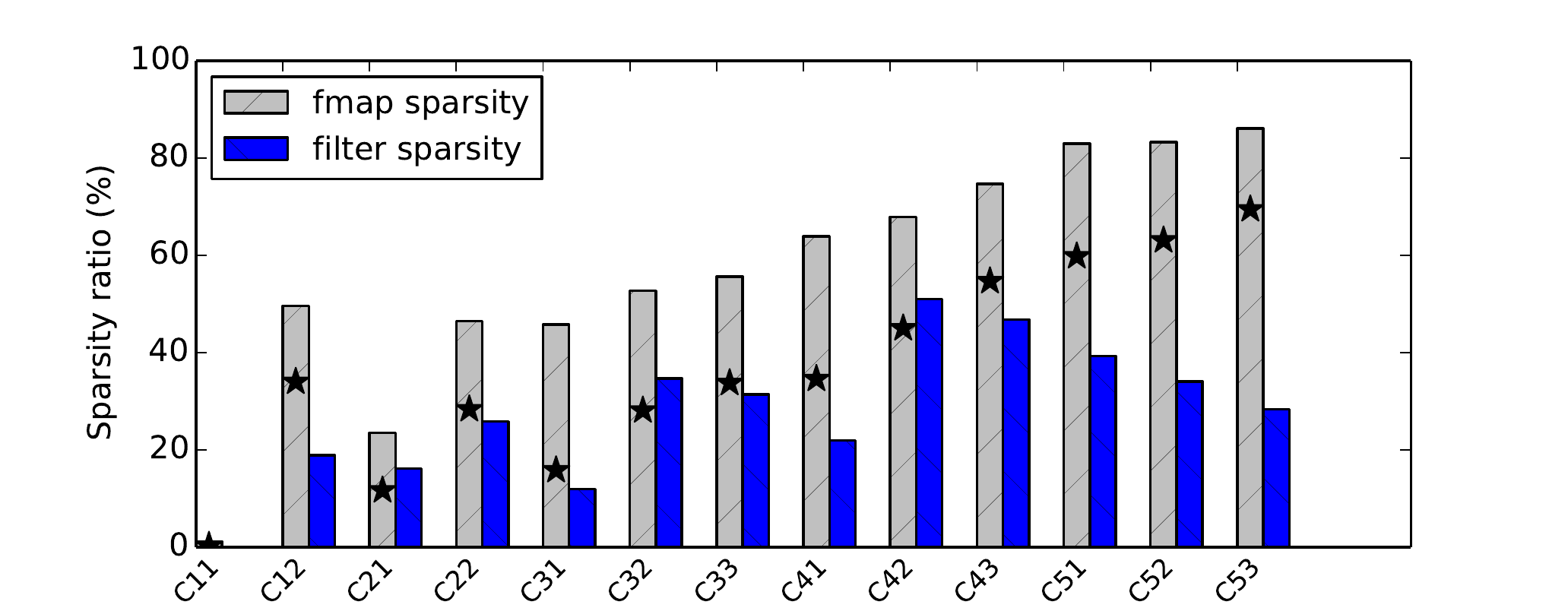}
                \caption{VGGNet.}
                
        \end{subfigure}%
        \hspace{-1.8em}%
        \begin{subfigure}[b]{0.50\textwidth}
        \centering
                \includegraphics[width=\textwidth]{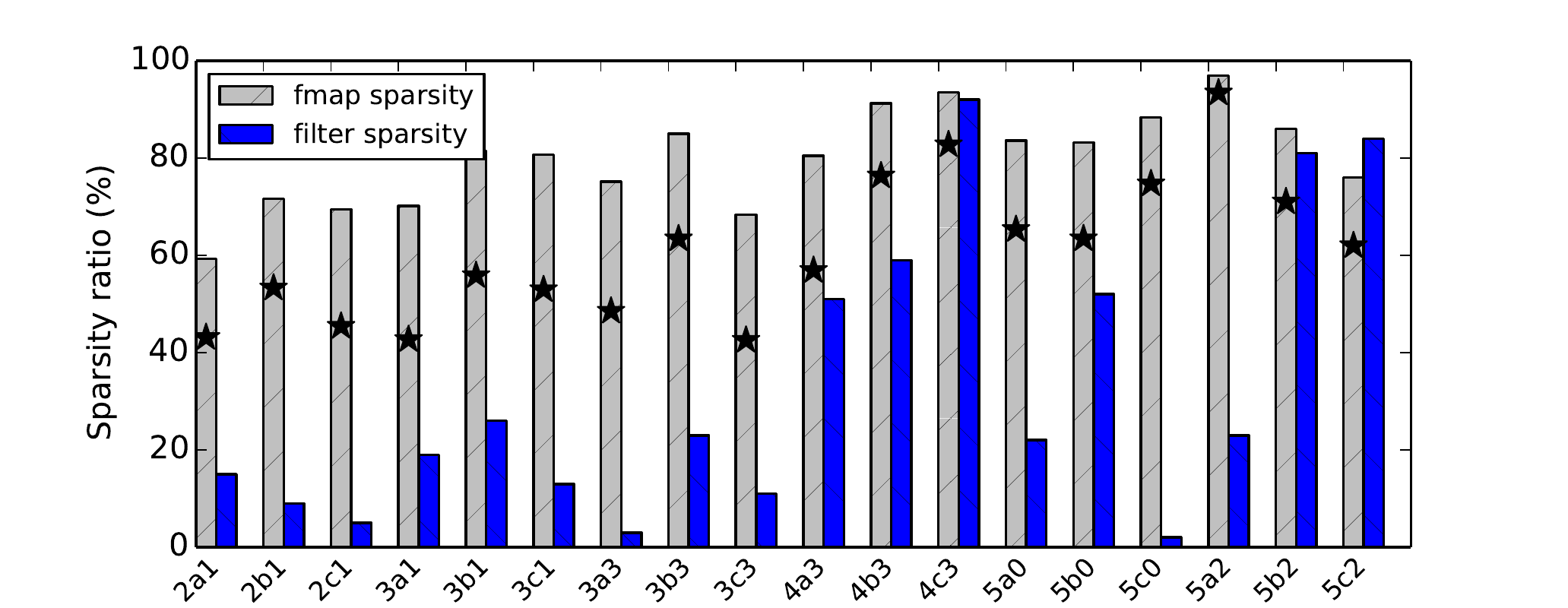}
                \caption{ResNet.}
                \label{fig:tiger}
        \end{subfigure}%
        \hspace{-1.8em}%
        \begin{subfigure}[b]{0.50\textwidth}
        \centering
                \includegraphics[width=\textwidth]{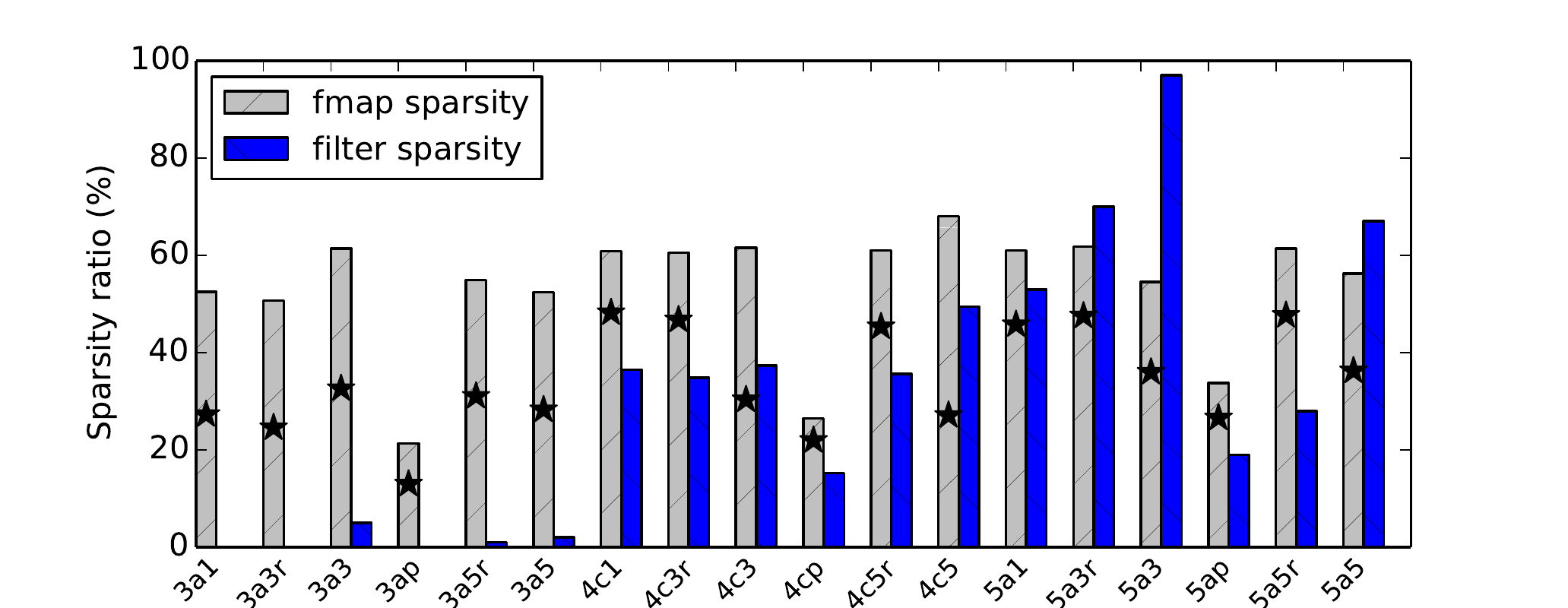}
                \caption{GoogleNet.}
                \label{fig:mouse}
        \end{subfigure}
        \caption{Sparsity in the filters and input feature maps for
AlexNet, VGGNet, ResNet, and GoogleNet. The $\star$ marker indicates
the percentage of zeros in the output of the \imcol transformation
that is skipped on-the-fly by our design.}
        \label{fig:sparsity}
\end{figure*}

We built a prototype of our design in Verilog and synthesized
it using Synopsys Design Compiler
with FreePDK 45nmcm technology~\cite{freepdk:45nm}. Our design achieves
a maximum of 500 MHz frequency. FreePDK 45 does not include SRAM
cells. Thus, we separately model the area and power of all the
SRAM/DRAM using Cacti 7.0~\cite{cacti7}. Table~\ref{table:spotsconfig}
provides the parameters of the SPOTS prototype and the area breakdown
for different components. We perform a cycle-accurate simulation
of the RTL model of SPOTS in Verilog using Verilator. We used the traces from our RTL simulation
and estimated the power consumption of our design with Synopsys's
PowerPrime tool. During our simulation, we executed each layer at a
time. The pruned weights are preprocessed and are provided in our
proposed sparse format. For the input feature map, we extracted each
layer's data from the models in Caffe. We also developed additional
infrastructure to perform fast design space exploration and to collect
statistics.

\textbf{CPUs, GPUs, and other ASICs used for our evaluation.} We
compare our prototypes with CPUs, GPUs, and other ASICs. We use Caffe
to evaluate various CNN architectures on a modern CPU and GPU. The
details of the CPU and GPU that we use for the evaluation is shown in
Table~\ref{table:config}. The CPU and GPU we used in our experiments
are manufactured with 22 nm and 16 nm cell technology, compared to 45
nm technology used for SPOTS.  The Caffe framework uses IntelMKL for
CPU computation and Nvidia's CUDA library, cuSparse, for the GPU
computation. Similar to our design, Caffe adopts a \imcol\texttt{+}
GEMM approach for doing convolution layers. We measured the energy
consumption of the XEON CPU using Processor Counter Monitor
(PCM)~\cite{intelpcm:power}. For GPU, we measured the power
consumption with NVIDIA System Management Interface
(Nvidia-smi)~\cite{nvidia:power} that queries the power using the
built-in sensors. According to NVIDIA, the reported data is accurate
(\ie, within $\pm$ 5 Watt).

\begin{table}[t]
  \centering
  \scriptsize
  \caption{\small Network characteristics and their the top1 and top5 result accuracy for the pruned and the original model. Weights and activations assume a data-type size of two bytes.}
  \begin{tabular}{|c|c|c|c|c|c|c|c|c|}
    \hline
    Model  & \#conv & Max. Layer & Max. & Median   & Median & Dataset & Baseline(\%)  &  Our Pruning (\%) \\ 
    ~      & Layer & Feature map & Layer & Kernel  & Stride &~ &Top1/Top5 Acc &  Top1/Top5 Acc \\
           &       &             & Weight & Size   & Size   &  &               &                \\ \hline
    AlexNet~\cite{alex:alexnet:2017} & 5 & 0.3 MB & 2.5 MB & 3 & 1 & Imagenet & 56.81/79.95 & 55.25/78.62  \\ \hline
    VGGNet~\cite{vggnet:arxiv} & 13  & 6.1 MB & 4.5 MB & 3 & 1 & Imagenet & 68.27/88.36 & 67.18/88.16 \\ \hline
    GoogleNet~\cite{Szegedy:googlenet:2015} & 57 &  0.36 MB & 1.3 MB & 1 & 1 & Imagenet & 68.92/89.14 & 66.22/87.53 \\ \hline
    ResNet~\cite{Kaiming:resnet:2016} & 53 & 1.5 MB & 4.5 MB & 1 & 1 & Imagenet & 77.71/90.66 & 69.71/89.30 \\ \hline
  \end{tabular}
  \label{table:accuracynetwork}
\end{table}

\begin{table}[t]
  \centering
  \scriptsize
  \caption{The CPU and GPU configurations.}
  \begin{tabular}{|c|c|c|c|c|c|}
    \hline
    Platform & Number of cores & Frequency & Main memory & Technology node & Cache \\ \hline

    Intel Xeon E5-V3 & 4 & 3 GHz & 32 GB DDR4  & 22nm & 10 MB   \\
                     &   &       &  (2666 Mhz) &      &  Smart Cache \\ \hline
        
    Titan X Pascal & 3584 & 1.53 GHz & 24 GB of GDDR5  & 16nm & - \\
                   &      &          &  (peak bandwdith 480 GB/S) & &  \\ \hline 
  \end{tabular}  
  \label{table:config}
  \vspace{-3mm}
\end{table}

\textbf{Eyeriss.} Although there are many prior ASICs on accelerating
CNN networks, most of them report relative performance numbers, and
their designs are not publicly available.  We use
Eyeriss~\cite{chen:eyeriss:journal:2017,chen:eyeriss:isca:2016}, which
is an ASIC designed for accelerating sparse CNNs, to compare against
our design. Eyeriss uses a row-stationary (RS) dataflow to maximize
data reuse and minimize expensive data movements. Further, data
compression and data gating techniques are applied to improve energy
efficiency. Eyeriss chip~\cite{chen:eyeriss:journal:2017} is composed
of 168 Processing Elements (PEs) structured as a $12\times14$
array. The PEs are connected with a network-on-chip (NOC) that enables
multicast and point-to-point single-cycle data delivery to support the
RS dataflow. We measure the performance of Eyeriss using the publicly
available simulator~\cite{Gao:tangram:asplos:2019}. Eyeriss chip is
fabricated at 65 nm CMOS and operates at 200 MHz clock
frequency. Since we used a different cell technology (\ie, 45 nm) for
SPOTS, we assume that the frequency of Eyeriss to be exactly equal to
the frequency of SPOTS when we report the execution time. We also
configured Eyeriss to use the same number of MAC units and on-chip
memory as SPOTS.  Additionally, both SPOTS and Eyeriss designs use
16-bit fixed-point inputs.

\textbf{Gemmini} is a recent open-source full-stack DNN accelerator
generator. Gemmini~\cite{gemmini:arxiv:2021} can be used to explore
the design-space of efficient DNN accelerators. The core unit in
Gemmini is a systolic array composed of processing elements
(PEs). Each PE can perform dot products and accumulations. The PEs
reads the data from local, explicitly managed scratch-pad of banked
SRAMs. Gemmini uses a two-level hierarchy, first composed of tiles,
where tiles are connected via explicit pipeline registers. Each tile
can be further broken down into an array of PEs. Most design
parameters can be adjusted to explore various designs. We failed to
build a design with an exact total number of PEs as SPOTS. Thus, we
used tiles with $32\times32$ PEs for Gemmini, which translates to a
total of 1024 MAC units, which is 2$\times$ more MAC units than our
prototype.  We also set the scratch-pad and banking size parameters in
Gemmini to match the on-chip memory used for SPOTS. Like our approach
for measuring the cycle counts, Gemmini's simulator reports the cycle
counts from simulating the design with Verilator.

\begin{figure}[t]
        \begin{subfigure}[b]{0.49\textwidth}
        \centering
                \includegraphics[width=\textwidth]{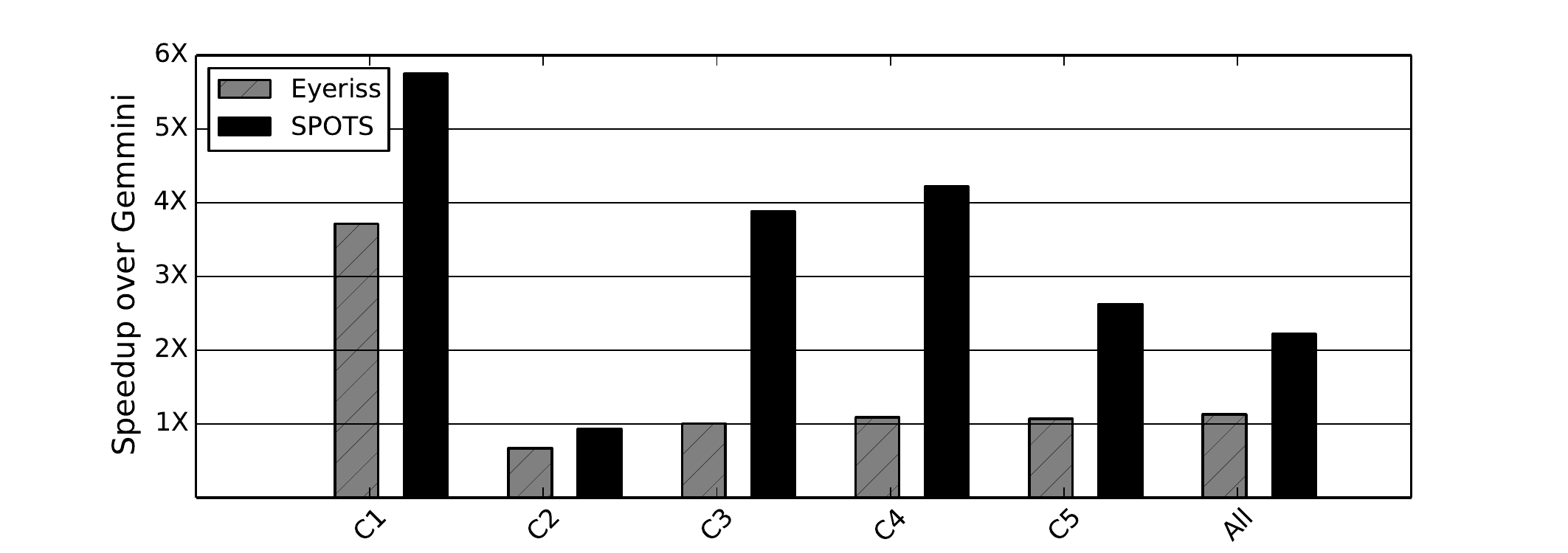}
                \caption{AlexNet}
                \label{fig:sub-first}
              
        \end{subfigure}%
        \hspace{-1.8em}%
        \begin{subfigure}[b]{0.49\textwidth}
        \centering
                \includegraphics[width=\textwidth]{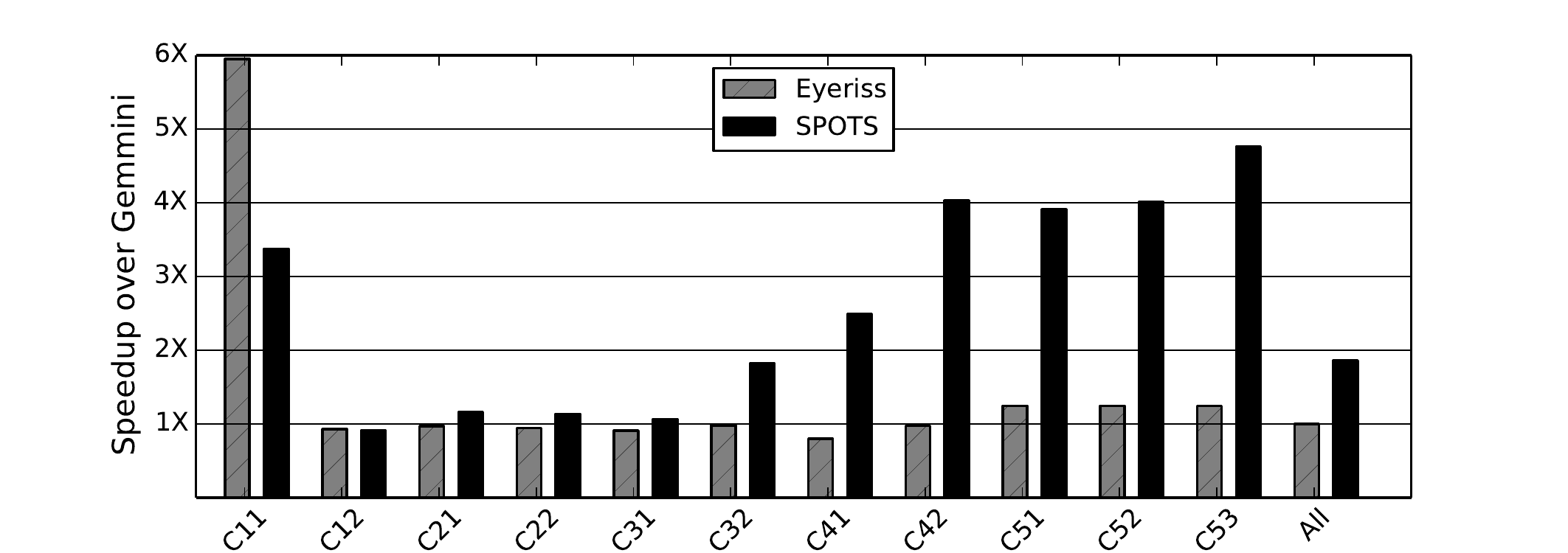}
                \caption{VGGNet}
                \label{fig:sub-second}
                
        \end{subfigure}%
        \hspace{-1.8em}%
        \begin{subfigure}[b]{0.49\textwidth}
        \centering
                \includegraphics[width=\textwidth]{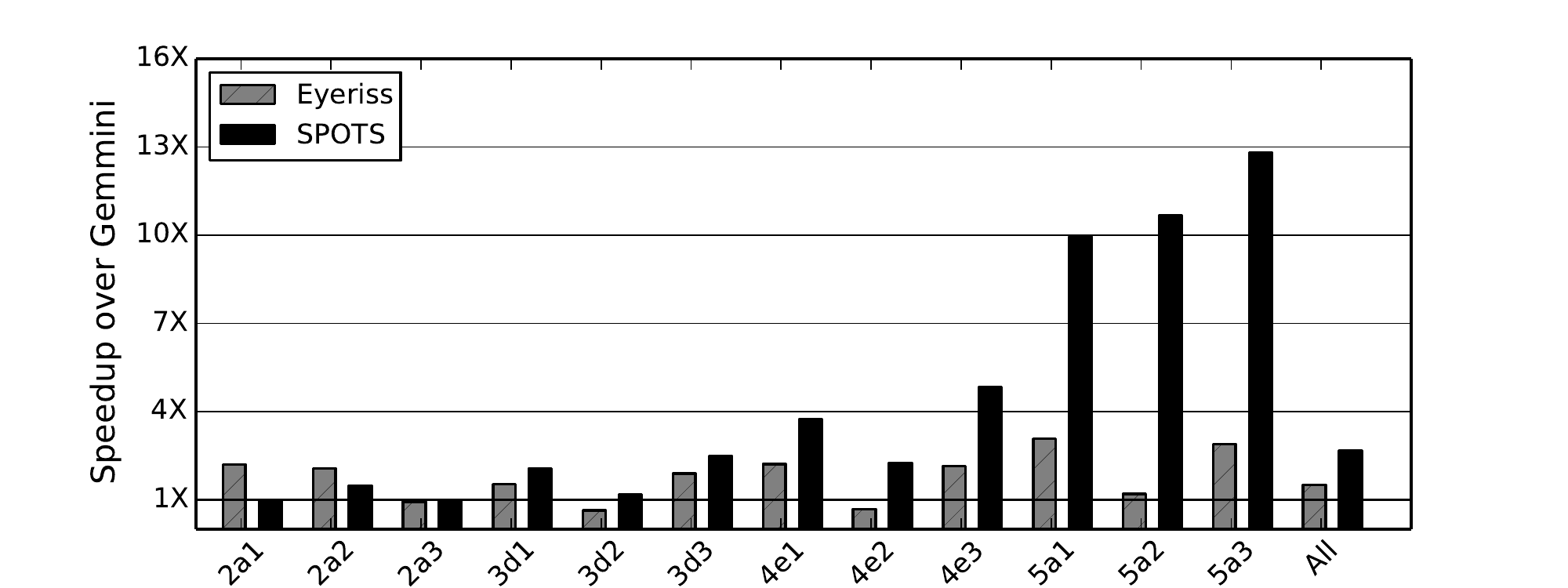}
                \caption{ResNet}
                \label{fig:sub-third}
        \end{subfigure}%
        \hspace{-1.8em}%
        \begin{subfigure}[b]{0.49\textwidth}
        \centering
                \includegraphics[width=\textwidth]{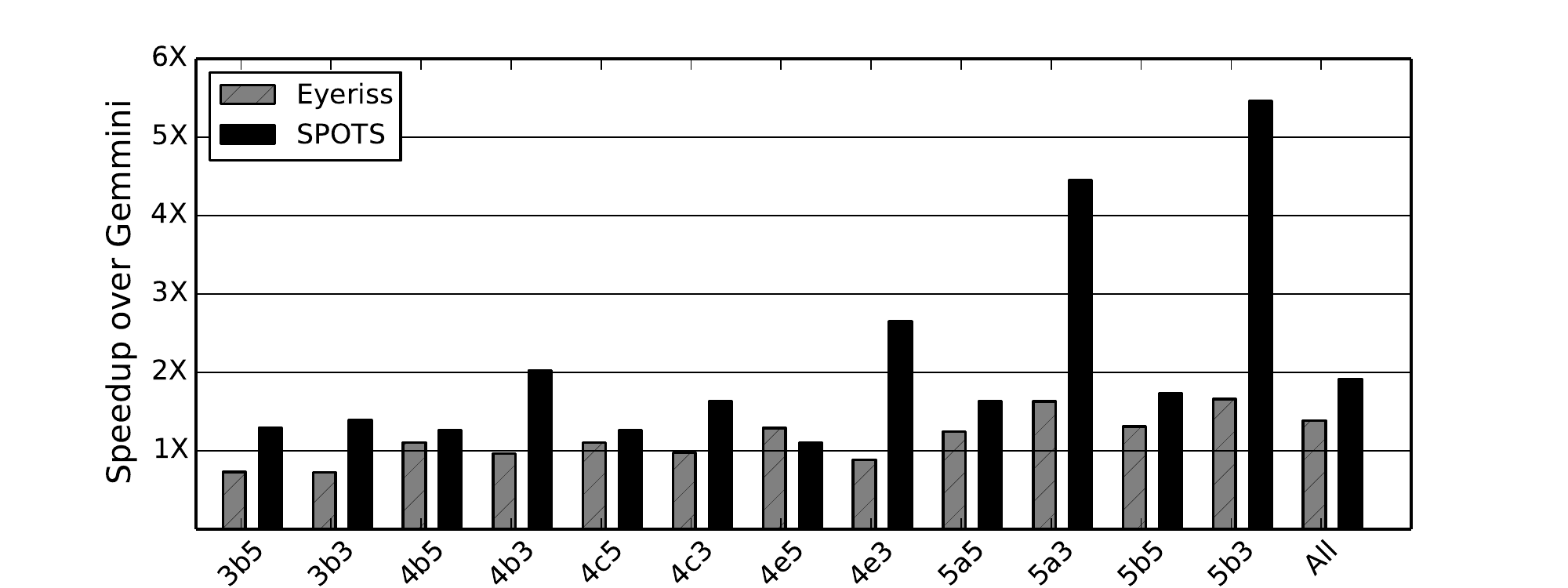}
                \caption{GoogleNet}
                \label{fig:sub-fourth}
        \end{subfigure}
        \caption{Speedup with SPOTS and Eyeriss over Gemmini for four CNNs: AlexNet,
VGGNet, ResNet, and GoogleNet. The figures show the speedup for selected layers from the top, middle, and bottom layers and the overall speedup (the last bar in each figure).}
        \label{fig:eyeriss-speedup}
\end{figure}

\textbf{CNN architectures and pruning.} We used four widely used CNN
architectures: AlexNet~\cite{alex:alexnet:2017},
VGGNet-16~\cite{vggnet:arxiv},
GoogleNet~\cite{Szegedy:googlenet:2015}, and
Resnet-50~\cite{Kaiming:resnet:2016} to evaluate our prototype. We
refer to VGGNet-16 and ResNet-50 as VGGNet and ResNet, respectively,
throughout the paper.  These four CNN architectures vary in the number
of layers, layer types, and sizes, as
Table~\ref{table:accuracynetwork} presents.  We used a batch size of
one for all of our experiments, which is the standard usage mode for
an inference task. We used the input images from the
Imagenet~\cite{imagenet}, a widely used dataset for image
classification tasks, to train the networks. We pruned all four
networks using the pruning algorithm based on Structure Sparsity
Learning~(SSL)~\cite{Wei:ssl:nips:2016}.  SSL is generic and can be
applied in different levels, including filters, channels, and
shapes. We applied SSL at the shape level. As our hardware exploits
sparsity at a much finer granularity than a shape, we optimize SSL by
pruning in a more fine-grained fashion. Specifically, we zeroed the
weights that are below the threshold in some but \textbf{not all} elements of a
shape. This generates zero blocks of a certain size (\ie, the number
of filters in the group). Figure~\ref{fig:pruningformat}(d) shows our
group-wise pruning. Figure~\ref{fig:sparsity} reports the sparsity in
the weights and input feature map after pruning for the layers of
various CNN architectures. It shows that sparsity varies across both
layers and networks. Finally, we retrained the pruned network to
regain its accuracy, which is the norm with pruning.
Table~\ref{table:accuracynetwork} reports the top-1 (\ie, the first
prediction is the correct result) and top-5 (\ie, the correct result
is in the first 5 predicted values) accuracies of the pruned network
and the original network. Our pruned networks are within 1\%-2\%
accuracy of the original model without pruning.

\section{Experimental Evaluation}
We demonstrate the performance and energy efficiency of SPOTS in
comparison to Eyeriss, Gemmini, CPU, and GPU implementations.
\begin{figure*}[t]
\begin{subfigure}[b]{0.49\textwidth}

  \centering  
 \includegraphics[width=\textwidth,angle=0]{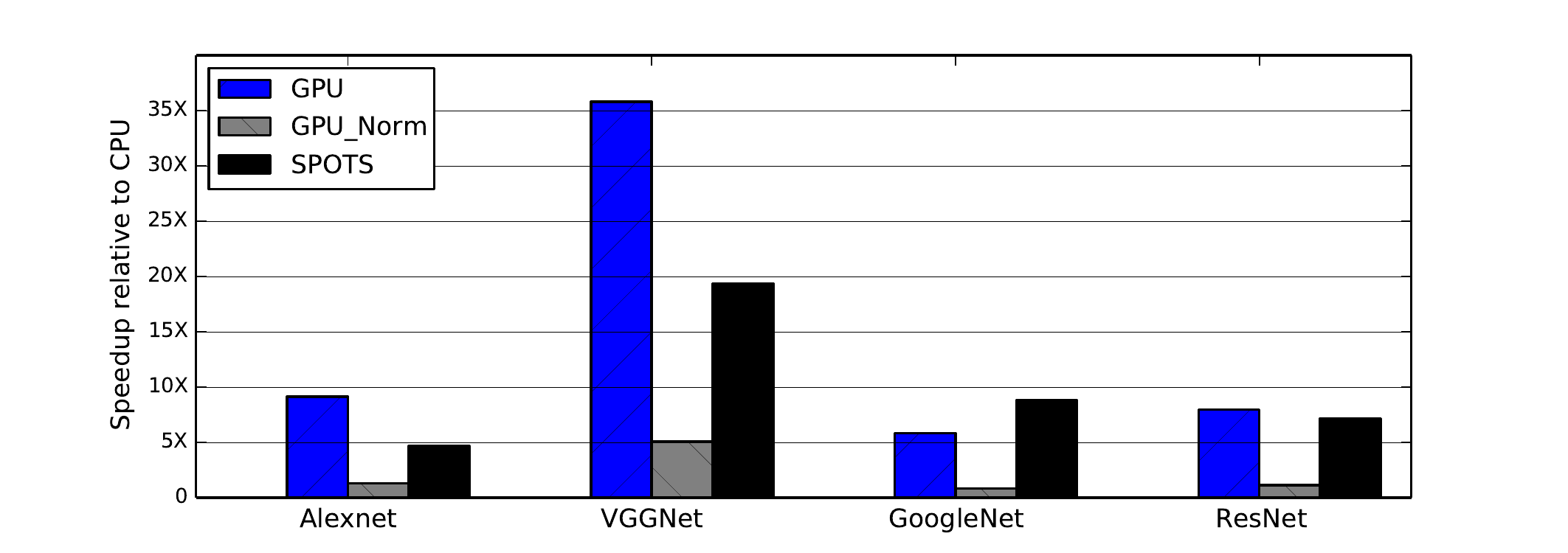}
 \caption{\small Speedup with SPOTS}
 \label{fig:cpuspeedup}
 \end{subfigure}%
 \hspace{-1.8em}%
\begin{subfigure}[b]{0.49\textwidth}
  \includegraphics[width=\textwidth,angle=0]{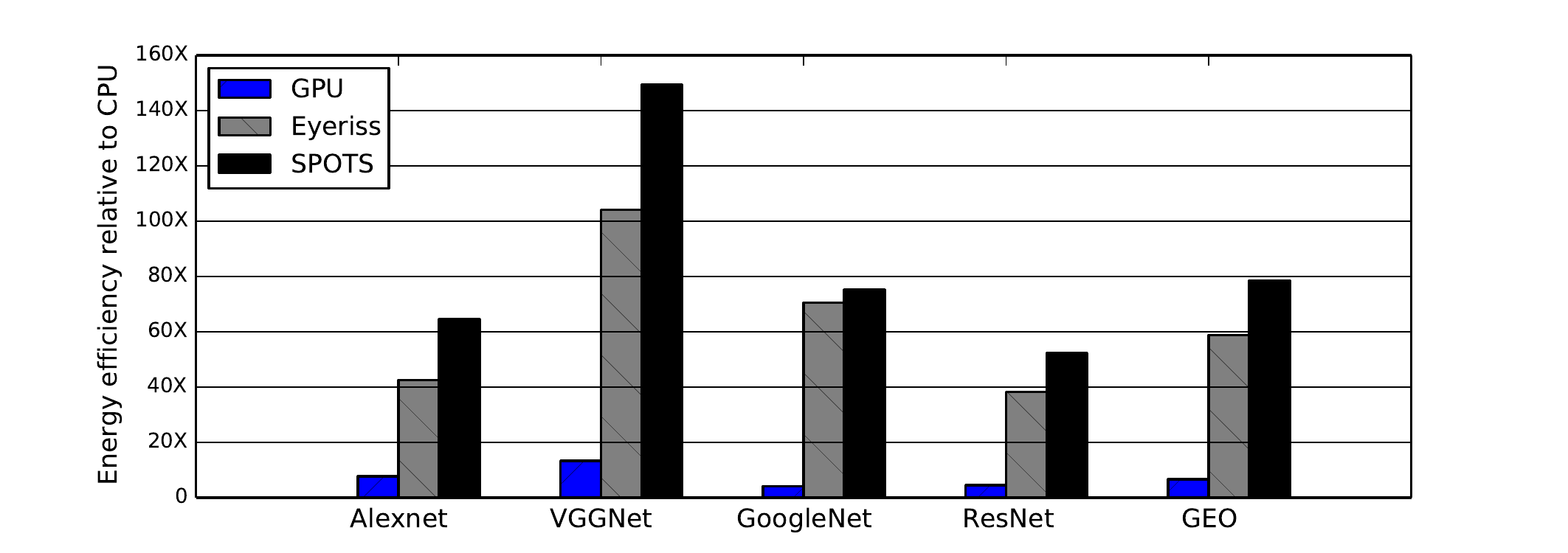}
  \caption{\small Energy efficiency with SPOTS}
\label{fig:energyefficiency}
\end{subfigure}

\caption{\small (a) Speedup with SPOTS, GPU, and GPU implementations with the normalized number of MAC units over the CPU implementation as the baseline. (b) The energy efficiency of SPOTS and GPU implementations compared to a CPU baseline.}

\end{figure*}

\textbf{Performance of SPOTS when compared to Eyeriss and Gemmini.}
Figure~\ref{fig:eyeriss-speedup} reports
the speedup of SPOTS and Eyeriss relative to Gemmini for all four CNN
architectures. The layers are
sorted for each CNN architecture based on where they appear in the
network (top, middle, or bottom).
Figure~\ref{fig:eyeriss-speedup}(a) reports the speedup for all layers
in AlexNet. On average, SPOTS is almost 2$\times$ faster than Eyeriss
and Gemmini. SPOTS is nearly $4\times$ faster than Eyeriss and Gemmini
for the layers in the middle, where the sparsity ratio in the two
inputs (\eg, weights and feature map) is higher. In addition to the
sparsity awareness that gives SPOTS an advantage over Eyeriss and
Gemmini, the layers in the middle and bottoms have more filters that
favor a tall systolic array.

Figure~\ref{fig:eyeriss-speedup}(b) reports the speedup for VGGNet. On
average, SPOTS is $1.85\times$ and $1.86\times$ faster than Eyeriss
and Gemmini, respectively. Similar to AlexNet, SPOTS achieves higher
speedup with layers with more sparsity. SPOTS perform slightly worse
than Eyeriss for the first two layers, where the number of filters is
relatively small, and the inputs are dense (See
Figure~{fig:sparsity}). Later in this section, we will demonstrate the
connection between the number of filters in a layer and the PE
utilization.
Figure~\ref{fig:eyeriss-speedup}(c) shows the speedup of SPOTS over
Eyeriss and Gemmini for ResNet. SPOTS is 1.77$\times$ and 2.66$\times$
faster than Eyeriss and Gemmini on average for ResNet. SPOTS is up to
8$\times$ and 13$\times$ faster than Eyeriss and Gemmini for the
layers where the weight and feature map sparsity is high.  Similar to
VGGNet, SPOTS performs slightly worse or similar to Eyeriss for the
first eight layers in ResNet because the first few layers in ResNet
have a few filters per each layer. Hence, PEs are underutilized
compared to layers in the middle or at the end of the network.
Figure~\ref{fig:eyeriss-speedup}(d) shows for GoogleNet, SPOTS is
1.38$\times$ and 1.91$\times$ faster than Eyeriss and Gemmini,
respectively. In contrast to other CNN architectures, GoogleNet has a
few convolutional layers at the beginning with a small number of
filters that do not favor our tall array. Thus, overall, SPOTS enjoys
less speedup for GoogleNet compared to the three other networks.

\textbf{Performance comparison with CPUs and GPUs.}  We evaluate the
performance and energy efficiency of SPOTS in comparison to execution
with CPUs and GPUs.  Figure~\ref{fig:cpuspeedup} reports the speedup
of SPOTS for the convolution layers over the CPU implementation. SPOTS
has 5$\times$, 20$\times$, 6$\times$ and 8$\times$ speedup over the
CPU implementations using Intel MKL for AlexNet, VGGNet, GoogleNet,
and ResNet, respectively. SPOTS attains this speedup while operating
at a frequency almost 6$\times$ less than the
CPU. Figure~\ref{fig:cpuspeedup} also shows the speedup of GPUs for
the convolution layers over the CPU implementation.  Compared to GPUs,
SPOTS is about 2$\times$ slower than GPU for AlexNet and VGGNet, while
performs slightly better or similar to GPU for GoogleNet and
ResNet. VGGNet and AlexNet layers are relatively larger than the other
two networks, resulting in larger matrices that favor the GPU with
abundant MAC units compared to SPOTS. The second bar
Figure~\ref{fig:cpuspeedup})(a) shows the GPU performance when its
number of MAC units is normalized to the number of MAC units in SPOTS.
For the normalized number of MAC units, SPOTS outperforms the GPU on
average by 6$\times$. Finally, some prior work observed that the
performance degrades for CPUs and GPUs when the sparse features are
used when the networks are pruned randomly. However, we observed that
using our structured pruning helps the CPU and GPU implementations to
attain higher overall performance using sparse linear algebra kernels.

\begin{figure*}[t]
        \begin{subfigure}[b]{0.49\textwidth}
        \centering
  		\includegraphics[width=\textwidth,angle=0]{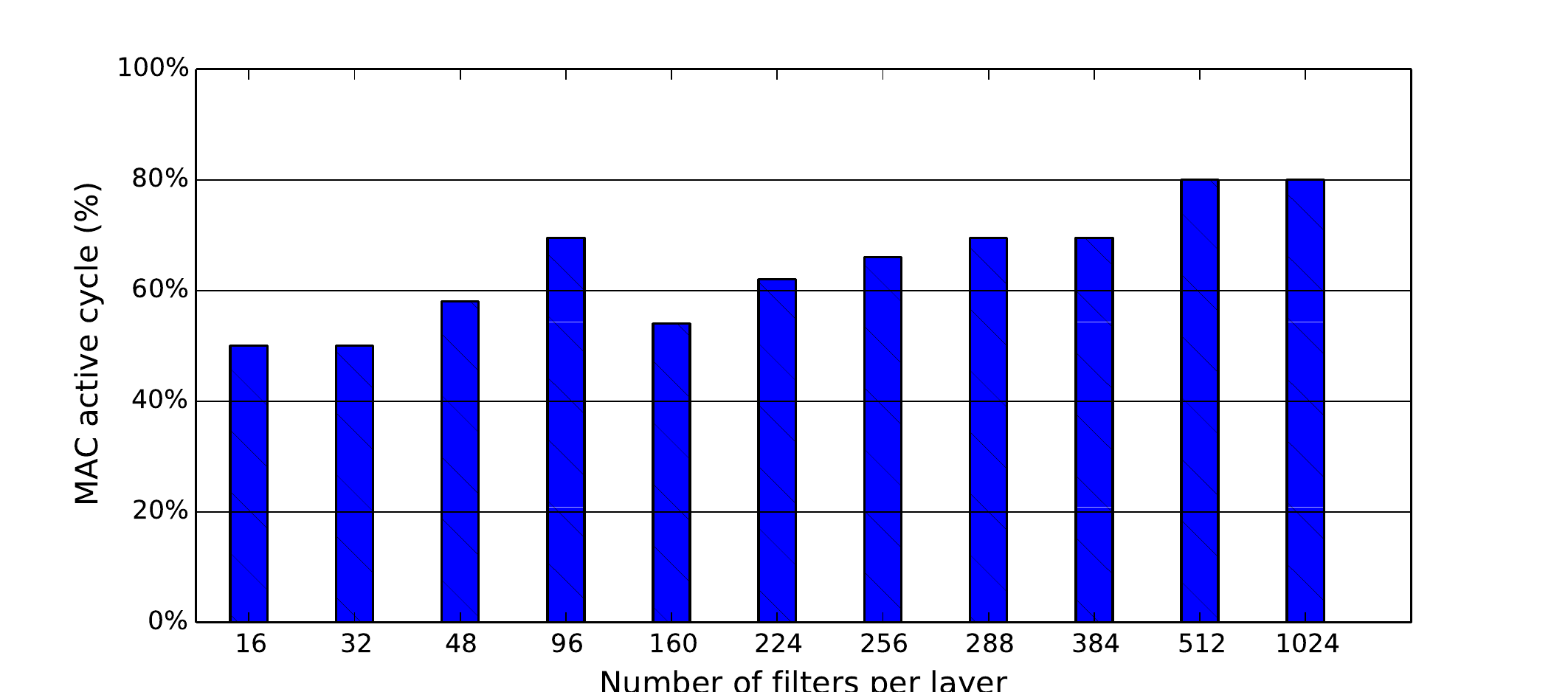}
		\caption{\small MAC utilization}
		\label{fig:macutil}
              
        \end{subfigure}%
        \hspace{-1.8em}%
        \begin{subfigure}[b]{0.49\textwidth}
        \centering
      \includegraphics[width=\textwidth,angle=0]{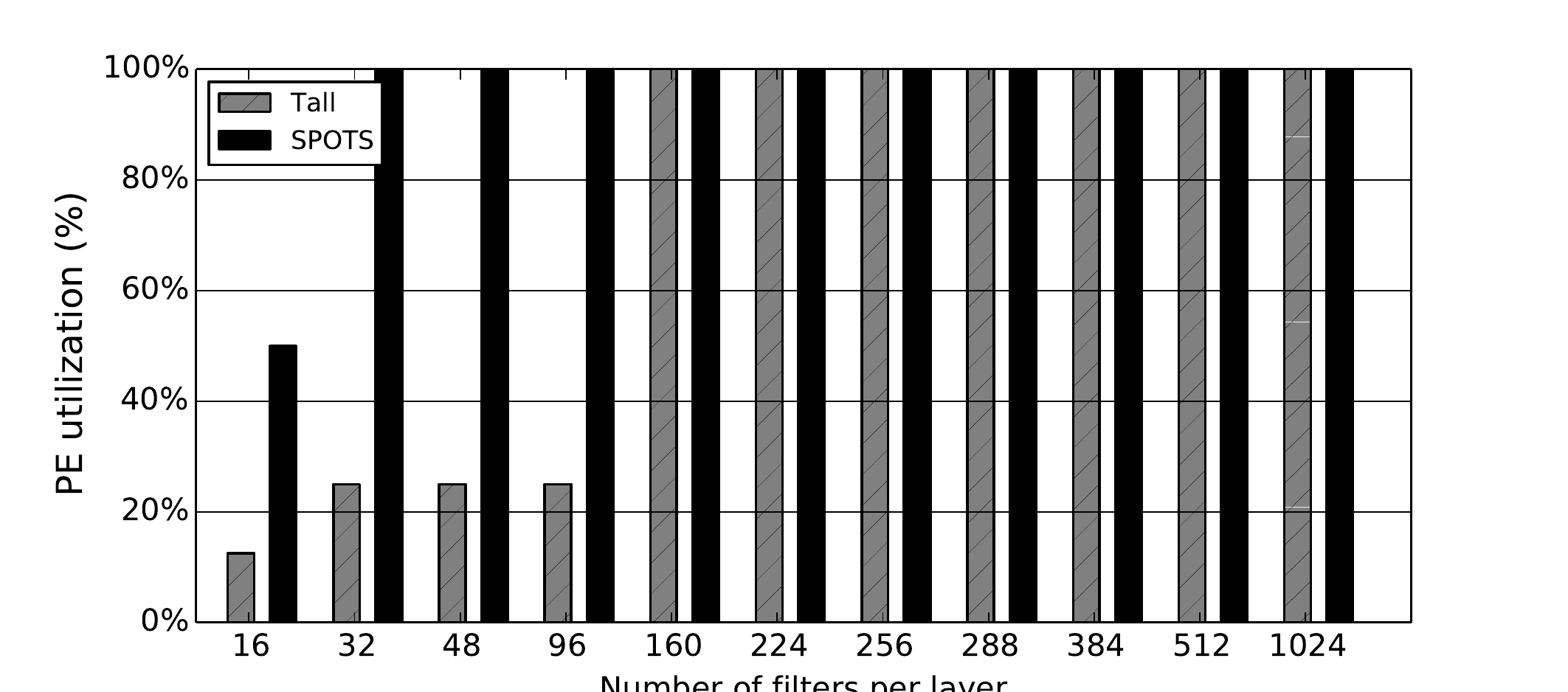}
    \caption{\small PE utilization}
    \label{fig:peutil}
              
        \end{subfigure}
        \caption{\small (a) MAC utilization (i.e., active cycles) for different
  filter sizes. (b) Comparing PE utilization of SPOTS and Tall systolic array (i.e., active PEs) for different filter sizes.}

\end{figure*}

\begin{figure*}[t]
        \begin{subfigure}[b]{0.49\textwidth}
        \centering
                \includegraphics[width=\textwidth]{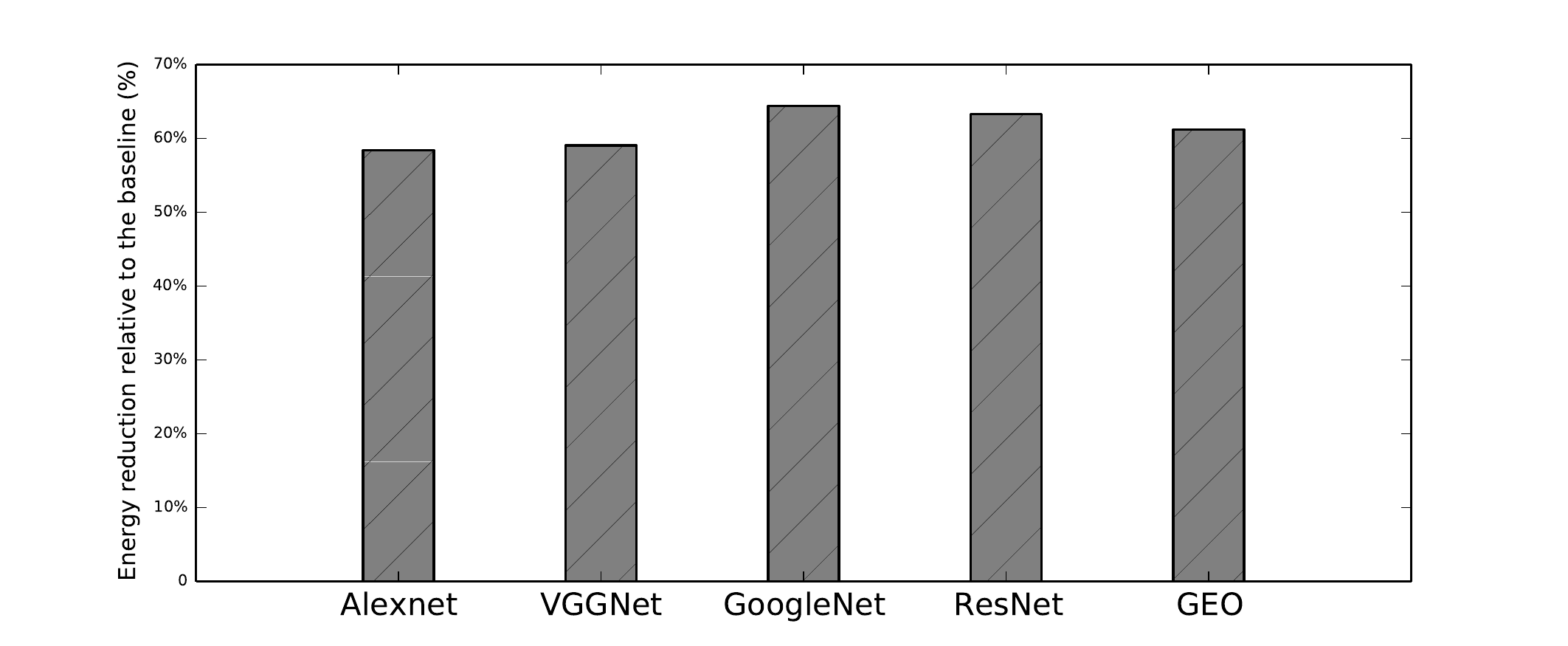}
                \caption{\small \imcol energy efficiency }
                \label{fig:im2col.efficiency}
        \end{subfigure}%
        \hspace{-1.8em}%
        \begin{subfigure}[b]{0.49\textwidth}
        \centering
                \includegraphics[width=\textwidth]{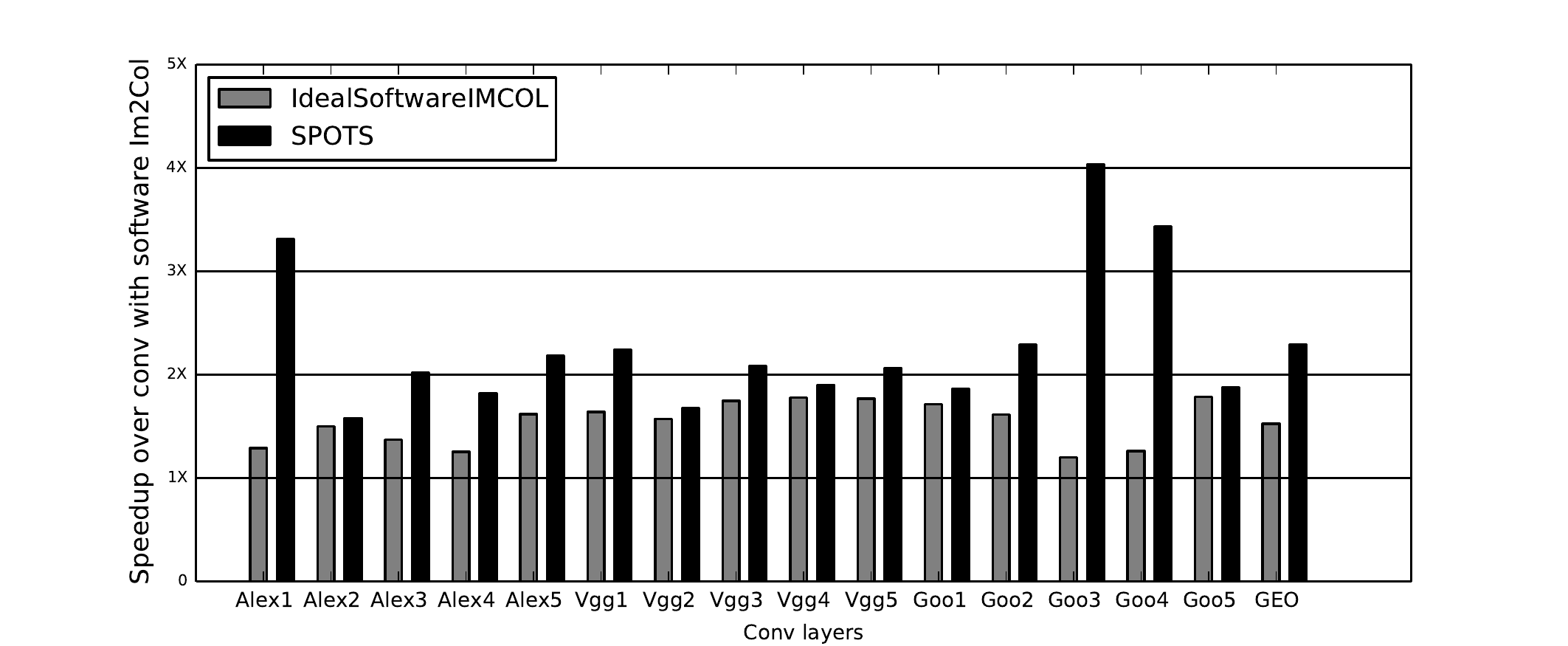}
                \caption{\small Hardware Vs. Software \imcol }
                \label{fig:im2colsoftware}
        \end{subfigure}%
        \hspace{-1.8em}
        \begin{subfigure}[b]{0.49\textwidth}
        \centering
  \includegraphics[width=\textwidth,angle=0]{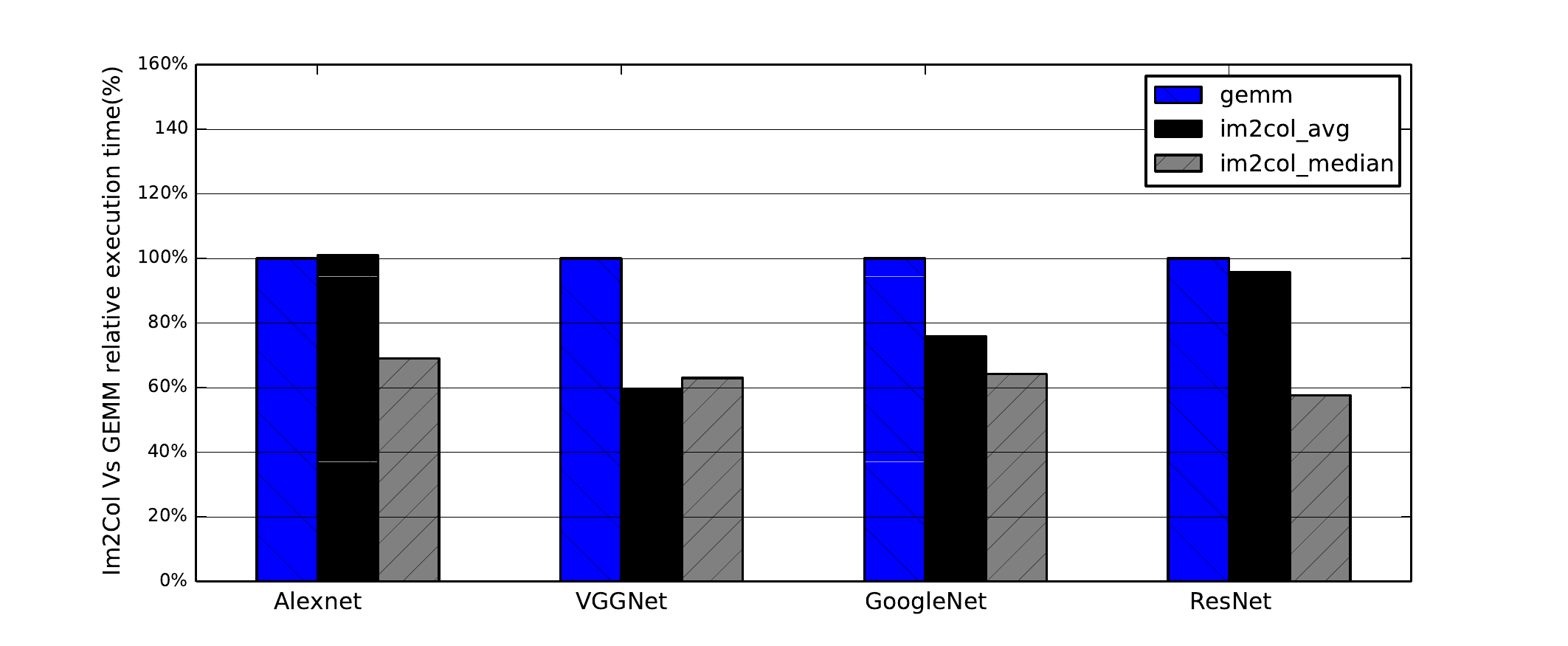}
\caption{\small \imcol Vs. GEMM}
\label{fig:im2colvsgemm}
                
        \end{subfigure}
        \caption{\small (a) The reduction in energy consumption for the \imcol
  unit of SPOTS over the baseline design. (b) Speedup with SPOTS over the software-based \imcol as the baseline \imcol design with no data reuse. (c) Fraction of the work performed by the \imcol unit when
  compared to GEMM (i.e., GEMM bar is 100\%). We report the average and
  the median for the \imcol's work.  When the mean exceeds the median,
  there will be instances where the \imcol does more work
  compared to GEMM for some layers.}

\end{figure*}

\textbf{Energy efficiency compared to CPUs and GPUs.}
Figure~\ref{fig:energyefficiency} demonstrates the energy efficiency
of SPOTS and GPU implementations when compared to a CPU baseline for
four CNNs. We did not include Gemmini energy results since their tool
does not report the power consumption.  The energy results include the
off-chip memory accesses in this data. Our accelerator consumes
78$\times$, 12$\times$, and 1.4$\times$ lesser energy than a CPU, a
GPU, and Eyeriss, respectively.

\textbf{Sensitivity to shapes of various layers.} Widely used CNN
networks vary in the depth and the number of filters used in each
layer. Even within a CNN, the layer shape and filter sizes can vary
significantly. The dynamic reconfigurability in SPOTS provides
flexibility to use the GEMM unit as a tall systolic array or as
multiple small systolic arrays, which allows it to adapt to various
shapes and filter sizes.
When the filters are small (\eg, less than 128), the GEMM unit is
configured as multiple small systolic arrays, which use different
\imcol units.  All the PEs in the systolic array are active 100\% of
the time for all filter sizes other than 16 (See
Figure~\ref{fig:peutil}). In contrast, a tall systolic array without
the enhancement we proposed in Section~\ref{sec:flexible} fails to
achieve full PE utilization for smaller filter sizes, as
Figure~\ref{fig:peutil} shows.  Figure~\ref{fig:macutil} shows the
utilization of the multiply-accumulate units in the PEs of the
systolic array (\ie, active cycles) when the layer has a specified
number of filters (\ie, x-axis reports the size of the filter).  When
the filter size increases, we assign more rows to a PE, which can
fetch up to four elements per read operation. Hence, there are more
opportunities to keep the multiply-accumulate units in the PE active
(\ie, almost 80\% active cycles).

\textbf{Amount of work performed by \imcol and GEMM units in SPOTS.}
As the \imcol and the GEMM units are pipelined in SPOTS, ideally, the
work done by the \imcol unit and the GEMM unit should be
balanced. Figure~\ref{fig:im2colvsgemm} shows the relative percentage
of cycles where the \imcol and GEMM units are active relative to the
GEMM unit for the four CNN architectures. As we report the active
cycles relative to the GEMM unit, the bar for the GEMM unit is 100\%.
The average work performed by the \imcol unit and the GEMM unit are
almost similar for AlexNet and ResNet (\ie, the work is balanced).  In
contrast, GEMM dominates the total work in VGGNet. This data suggests
that adding more PE's to the GEMM unit may improve the overall
execution time for VGGNet.  As the \imcol unit is inactive due to low
bandwidth with AlexNet and ResNet, adding more PEs without increasing
the bandwidth will not improve performance.

\textbf{Energy efficiency from data reuse in the \imcol unit.} One of
the key idea in the \imcol's patch unit is to read the input feature
map only once from the SRAM and reuse the data with the help of local
buffers.  Figure~\ref{fig:im2col.efficiency} reports the percentage
decrease in energy consumed by using local buffers to reuse the data
in the patch units compared to a naive version of \imcol that access
SRAMs multiple times without any data reuse. On average, the
mechanisms that we added to reuse the input feature map in the patch
units result in the \imcol unit consuming 60\% less energy when
compared to the \imcol unit without such reuse.

\textbf{Comparing SPOTS with software \imcol.} SPOTS has a hardware
\imcol unit that performs the \imcol transformation on-the-fly.
Figure~\ref{fig:im2colsoftware} compares the speedup of using a
hardware \imcol unit compared to a software-based \imcol as the
baseline.  For the baseline system, the hardware only performs GEMM
while the CPU executes the \imcol. The figure also shows the ideal
situation for a software-based \imcol design where the software \imcol
and the hardware GEMM computations are overlapped. Even when we
provide an ideal scenario for the software \imcol, SPOTS outperforms the
ideal software-based \imcol. On average, SPOTS outperforms the
baseline software \imcol by 2.3$\times$, which shows the benefits of
our hardware \imcol unit.

\section{Related work}
\label{sec:related}
\begin{table*}
  \begin{small}
  \centering
   \caption{Qualitative comparison of SPOTS with prior work.} 
  \begin{tabular}{|c|c|c|c|c|c|c|c|c|} 
    \hline
    Accelerator & \multicolumn{4}{|c|}{Supports sparsity} & Supports & Adaptive to   \\\cline{2-5}
    & Feature map & Weight & Gate zero & Skip zero & pruned network & different layer \\
    &              &        &           &          &                &  shapes\\     
    \hline
    \textbf{Eyeriss~\cite{chen:eyeriss:journal:2017}} & \cmark & \xmark & \cmark & \xmark & \xmark & \xmark  \\ \hline    
    \textbf{Cnvlutin~\cite{Albericio:Cnvlutin}} & \cmark & \xmark & \cmark & \cmark & \xmark & \xmark \\ \hline    
    \textbf{CambriconS~\cite{zhou:cambricons:micro:2018}} & \cmark & \cmark & \xmark & \cmark & \cmark (structured) & \xmark \\ \hline    
    \textbf{SCNN~\cite{parashar:scnn:isca:2017}} & \cmark & \cmark & \cmark & \cmark & \cmark(random) & \xmark \\ \hline 
    \textbf{CMSA~\cite{Xu:reconfig:cnn:TACO:2021}} & \xmark & \xmark & \xmark & \xmark & \xmark (random) & \cmark \\ \hline

    \textbf{Column comb.~\cite{kung:packingsystolic:asplos:2019}} & \xmark & \cmark & \xmark & \cmark & \cmark (structured) & \xmark \\ \hline    
    \textbf{SIGMA~\cite{qin:sigma:hpca:2020}} & \cmark & \cmark & \xmark & \cmark & \cmark (random) & \cmark \\ \hline 

    \textbf{SPOTS (this work)} & \textbf{\cmark} & \textbf{\cmark} & \textbf{\cmark} & \textbf{\cmark} & \textbf{\cmark structured} & \textbf{\cmark} \\ \hline
  \end{tabular}
    \label{table:qualcompare}
      \end{small}

\end{table*}

There is a large body of literature on using custom hardware
accelerators to improve the performance and energy efficiency of
neural
networks~\cite{Albericio:Cnvlutin,zhang:cambriconx:micro:2016,parashar:scnn:isca:2017,chen:eyeriss:journal:2017,Reagen:Minerva:isca:2016,chen:DaDianNao:micro:2014,han:eie:deep:isca:2016,sharify:laconic:isca:2019,Albericio:Bit-Pragmatic:micro,chen:DaDianNao:micro:2014,sharma:DNNWEAVER:micro:2016,huang:ecnn:micro:2019,Chunhua:GoSPA:isca:2021,gemmini:arxiv:2021}.
Table~\ref{table:qualcompare} qualitatively compares SPOTS with more
closely related work. The table shows that SPOTS supports various
operations in CNNs, is adaptive to different layers' shapes to keep
the PE's utilization high for different scenarios, and efficiently
handles sparsity in both feature map and weights.

\textbf{Support for sparse inputs.} Prior work has improved energy
efficiency by supporting sparse inputs during
inference. Cnvlutin~\cite{Albericio:Cnvlutin} exploits sparsity in the
input feature map to skip multiplication operations and to avoid data
movement with zero elements.
CambriconX~\cite{zhang:cambriconx:micro:2016} supports sparsity in the
weights.  Similar to our work, SCNN~\cite{parashar:scnn:isca:2017} and
CambriconS~\cite{zhou:cambricons:micro:2018} support sparsity in both
the feature map and the weights to improve energy efficiency and
performance.
Prior work also uses data gating techniques to reduce the power
consumption when the operands are
zeros~\cite{Reagen:Minerva:isca:2016,chen:eyeriss:journal:2017}. While
this technique is effective in reducing power consumption, it does not
reduce the number of effective operations. Similar to SPOTS, prior
hardware designs have developed techniques to skip zeros and
to minimize data
transfer~\cite{han:eie:deep:isca:2016,parashar:scnn:isca:2017,Huiyu:tfe:micro:2020}.

\textbf{Support for various layers in CNNs.}  Often,
hardware designs are customized for one type of computation and do not
support all types of layers in CNNs, such as pooling
layers~\cite{kung:packingsystolic:asplos:2019,asghri:ERIDANUS:micro}.
EIE~\cite{han:eie:deep:isca:2016} is intended for the fully connected
layers in CNNs. It stores the input feature map and filters in a
compressed format and passes only non-zero operands to the
multipliers. In contrast, SCNN~\cite{parashar:scnn:isca:2017} and
Eyeriss~\cite{chen:eyeriss:journal:2017,chen:eyeriss:isca:2016}
primarily focus on the convolution layers. Hence, they can
underperform for the fully-connected layers. SCNN can achieve 25\% of
peak throughput when performing the fully connected CNN
layers. Similarly, Eyeriss provides significant energy gains only when
batch sizes are larger than 16. In contrast, SPOTS supports all the common layers that exist in CNNs.

\textbf{Systolic array designs for CNNs.}  Recent
work~\cite{kung:packingsystolic:asplos:2019,sparsetpu:ics:2020} uses a
preprocessing step (i.e., column combining) to pack a sparse CNN into
a denser form before passing the inputs to a systolic array for
GEMM. It is unclear how to prepare input feature maps for matrix
multiplication. It will not provide benefits when there is abundant
sparsity in the input feature map. Our group-wise pruning provides
higher accuracy than the column combining method.  Simultaneous
multithreaded systolic array (SMT-SA)~\cite{Shomron:SMT-SA:2019}
addresses the underutilization and load imbalance introduced by
random pruning of the weights in a CNN. Similar to SPOTS, recent
work~\cite{liu:sparsegemm:letter:2020} utilizes a structured pruning
accompanied with a novel data format called density-bound block (DBB)
better to map the sparse inputs to the systolic architecture. Similar to SPOTS, Gemmini~\cite{gemmini:arxiv:2021} uses a GEMM to accelerate CNNs. The authors explored both software and hardware \imcol units. Similar to our work, their results demonstrate that using a hardware \imcol can significantly improve performance.  Unlike SPOTS, Gemmini is not sparse-aware. In addition, the PE structured in their design is rigid, resulting in PEs underutilization for certain layer shapes.  

\textbf{Flexible interconnects.} Flexible interconnects between PEs
are useful in supporting various filter
sizes~\cite{kwon:maeri:asplos:2018,qin:sigma:hpca:2020}.
Maeri~\cite{kwon:maeri:asplos:2018} enables a flexible dataflow
mapping over DNN accelerators using a tree-based reconfigurable
interconnects network. A downside of MAERI is that it does not handle
input feature map sparsity. Similarly,
FlexFlow~\cite{lu:flexflow:hpca:2017} develops a flexible dataflow
architecture that exploits different types of parallelism along with
different CNN workloads. In contrast to them, SPOTS uses a regular
interconnect network between the PEs.
SIGMA~\cite{qin:sigma:hpca:2020} is another recent work that proposes
a flexible non-blocking interconnect to achieve high compute
utilization across layers of varying shapes.  SIGMA is primarily
optimized for high-precision inputs during the training
phase. Besides, they solely focus on the GEMM and do not study the
\imcol transformation and support other types of layers in a CNN. Recent work ~\cite{Xu:reconfig:cnn:TACO:2021} design a configurable multi-directional systolic array (CMSA) that improves the PE utilization for small-scale convolution or depthwise convolution. However, their design solely focuses on improving the PE's utilization and thus does not address other aspects such as sparse inputs and \imcol design. 

\textbf{Other accelerators.} Recent work has explored the design space
of hardware accelerators for optimal dataflow and scheduling schemes
for neural
networks~\cite{kwon:maestro:micro:2019,TETRIS:Mingyu,Parashar:timeloop:ispass:2019,yang:Interstellar:asplos:2020,Gao:tangram:asplos:2019}.
Beyond sparse convolution, custom hardware to accelerate sparse
matrix-matrix multiplication with very sparse matrices (\ie, density
below 1\%) have also been
explored~\cite{Hegde:extensor:micro,Hojbar:SPAGHETTI:hpca:2021,pal:outerspace,sparch:Zhang:hpca:2020,soltaniyeh2020synergistic,mapraptor:Srivastava:micro:2020}.
Tensor Processing Unit (TPU)~\cite{tpu:google:isca} is an ASIC that
has matrix multiplication as its core computation block to accelerate
CNNs. TPU requires the host CPU to perform some data
reorganization and does not support sparse inputs.

\section{Conclusion}
This paper proposes SPOTS, a hardware accelerator for sparse CNNs with a matrix multiplication formulation of convolution using the \imcol
transformation. The hardware \imcol unit reads the input feature map
only once, reuses the data, and executes in parallel with a tall
systolic array for the GEMM unit. We add flexibility to the systolic array that allows it to achieve high PE utilization for CNN layers of varying sizes and shapes. SPOTS supports sparsity both in the input feature map and the filters. SPOTS is faster and more energy efficient than state-of-the-art systolic array-based ASICs, CPU, and GPU implementations for sparse CNNs.

\begin{acks}                            
 This material is based upon work supported in part by the
 \grantsponsor{GS100000001}{National Science
   Foundation}{http://dx.doi.org/10.13039/100000001} under Grant
 No.~\grantnum{GS100000001}{1908798}.
  Any opinions, findings, and conclusions or recommendations expressed
  in this material are those of the authors and do not necessarily
  reflect the views of the National Science Foundation.
\end{acks}

\bibliographystyle{ACM-Reference-Format}
\bibliography{cnn}

\end{document}